\documentclass[12pt,draftclsnofoot,onecolumn]{IEEEtran}
\usepackage{amsmath,amssymb,amsmath,amsfonts}
\usepackage{bm}
\usepackage{bbm}
\usepackage{graphicx}
\usepackage{cite}
\usepackage{epstopdf}
\usepackage{balance}
\usepackage{gensymb}
\usepackage{color}
\usepackage{lipsum}
\usepackage{float}
\usepackage{epstopdf}
\usepackage[mathcal]{eucal}
\usepackage{subfiles}
\usepackage[T1]{fontenc}
\usepackage{siunitx}
\usepackage[table]{xcolor}
\usepackage{tabulary}

\usepackage{epsfig,graphics,graphicx,amssymb,amstext,amsmath,algorithm,algorithmic,wrapfig,multirow}
\usepackage{array}
\usepackage{adjustbox}
\usepackage{cite}
\usepackage{times}
\usepackage{placeins}
\usepackage{textcomp}
\usepackage[font=small]{caption}
\usepackage{flushend}
\usepackage{xcolor}
\usepackage{tabularx}
\usepackage{bm}
\usepackage{enumerate}
\usepackage{tikz}
\usepackage{pgfplots}
\usepackage{epstopdf}
\usetikzlibrary{shapes,arrows}
\usepackage[utf8]{inputenc}
\usepackage{mathtools}
\usepackage[utf8]{inputenc}
\usepackage{xcolor}
\usepackage{tikz}
 \usepackage{comment}
\usetikzlibrary{chains,arrows,calc,positioning}

\pgfplotsset{compat=1.16}
\definecolor{bittersweet}{rgb}{1.0, 0.44, 0.37}
\definecolor{glaucous}{rgb}{0.38, 0.51, 0.71}
\definecolor{gainsboro}{rgb}{0.86, 0.86, 0.86}
\definecolor{babyblueeyes}{rgb}{0.63, 0.79, 0.95}
\definecolor{silver}{rgb}{0.75, 0.75, 0.75}
\definecolor{neoncarrot}{rgb}{1.0, 0.64, 0.26}

\usepackage{soul}

\usepackage{breqn}
\usepackage{booktabs}
\usepackage{multirow}

\usepackage{enumitem}
\usepackage{tabulary}

\usepackage[normalem]{ulem}
\usepackage[acronyms,nonumberlist,nopostdot,nomain,nogroupskip]{glossaries}
\newacronym{quic}{QUIC}{Quick UDP Internet Connections}
\newacronym{3gpp}{3GPP}{3rd Generation Partnership Project}
\newacronym{adc}{ADC}{Analog to Digital Converter}
\newacronym{5g}{5G}{5th generation}
\newacronym{aimd}{AIMD}{Additive Increase Multiplicative Decrease}
\newacronym{am}{AM}{Acknowledged Mode}
\newacronym{amc}{AMC}{Adaptive Modulation and Coding}
\newacronym{aqm}{AQM}{Active Queue Management}
\newacronym{awgn}{AGWN}{Additive White Gaussian Noise}
\newacronym{afd}{AFD}{Austin Fire Department}
\newacronym{balia}{BALIA}{Balanced Link Adaptation}
\newacronym{bdp}{BDP}{Bandwidth-Delay Product}
\newacronym{bf}{BF}{Beamforming}
\newacronym{cc}{CC}{Congestion Control}
\newacronym{cdf}{CDF}{Cumulative Distribution Function}
\newacronym{cn}{CN}{Core Network}
\newacronym{cqi}{CQI}{Channel Quality Information}
\newacronym{cp}{CP}{Control Plane}
\newacronym{csirs}{CSI-RS}{Channel State Information - Reference Signal}
\newacronym{dc}{DC}{Dual Connectivity}
\newacronym{dce}{DCE}{Direct Code Execution}
\newacronym{dci}{DCI}{Downlink Control Information}
\newacronym{dl}{DL}{Downlink}
\newacronym{dmr}{DMR}{Deadline Miss Ratio}
\newacronym{dmrs}{DMRS}{DeModulation Reference Signal}
\newacronym{e2e}{E2E}{End-to-End}
\newacronym{ecn}{ECN}{Explicit Congestion Notification}
\newacronym{edf}{EDF}{Earliest Deadline First}
\newacronym{enb}{eNB}{evolved Node Base}
\newacronym{epc}{EPC}{Evolved Packet Core}
\newacronym{es}{ES}{Edge Server}
\newacronym{fdma}{FDMA}{Frequency Division Multiple Access}
\newacronym{fdd}{FDD}{Frequency Division Duplexing}
\newacronym[firstplural=Radio Access Technologies (RATs)]{rat}{RAT}{Radio Access Technology}
\newacronym{fs}{FS}{Fast Switching}
\newacronym{ftp}{FTP}{File Transfer Protocol}
\newacronym{gnb}{gNB}{Next Generation Node Base}
\newacronym{harq}{HARQ}{Hybrid Automatic Repeat reQuest}
\newacronym{hetnet}{HetNet}{Heterogeneous Network}
\newacronym{hh}{HH}{Hard Handover}
\newacronym{hol}{HOL}{Head-of-Line}
\newacronym{ia}{IA}{Initial Access}
\newacronym{imt}{IMT}{International Mobile Telecommunication}
\newacronym{iot}{IoT}{Internet of Things}
\newacronym{los}{LOS}{line-of-sight}
\newacronym{lte}{LTE}{Long Term Evolution}
\newacronym{m2m}{M2M}{Machine to Machine}
\newacronym{mac}{MAC}{Medium Access Control}
\newacronym{mc}{MC}{Multi-Connectivity}
\newacronym{mcs}{MCS}{Modulation and Coding Scheme}
\newacronym{mec}{MEC}{Mobile Edge Cloud}
\newacronym{mi}{MI}{Mutual Information}
\newacronym{mimo}{MIMO}{Multiple Input, Multiple Output}
\newacronym{mmwave}{mmWave}{millimeter wave}
\newacronym{mr}{MR}{Maximum Rate}
\newacronym{mss}{MSS}{Maximum Segment Size}
\newacronym{mtd}{MTD}{Machine-Type Device}
\newacronym{mtu}{MTU}{Maximum Transmission Unit}
\newacronym{nfv}{NFV}{Network Function Virtualization}
\newacronym{nlos}{NLOS}{Non Line of Sight}
\newacronym{nr}{NR}{New Radio}
\newacronym{ofdm}{OFDM}{Orthogonal Frequency Division Multiplexing}
\newacronym{pdcch}{PDCCH}{Physical Downlonk Control Channel}
\newacronym{pdcp}{PDCP}{Packet Data Convergence Protocol}
\newacronym{pdsch}{PDSCH}{Physical Downlink Shared Channel}
\newacronym{pdu}{PDU}{Packet Data Unit}
\newacronym{pf}{PF}{Proportional Fair}
\newacronym{pgw}{PGW}{Packet Gateway}
\newacronym{phy}{PHY}{Physical}
\newacronym{pbch}{PBCH}{Physical Broadcast Channel}
\newacronym[plural=\gls{mme}s,firstplural=Mobility Management Entities (MMEs)]{mme}{MME}{Mobility Management Entity}
\newacronym{prb}{PRB}{Physical Resource Block}
\newacronym{pss}{PSS}{Primary Synchronization Signal}
\newacronym{pucch}{PUCCH}{Physical Uplink Control Channel}
\newacronym{pusch}{PUSCH}{Physical Uplink Shared Channel}
\newacronym{rach}{RACH}{Random Access Channel}
\newacronym{ran}{RAN}{Radio Access Network}
\newacronym{red}{RED}{Robotics Emergency Deployment}
\newacronym{rf}{RF}{Radio Frequency}
\newacronym{rlc}{RLC}{Radio Link Control}
\newacronym{rlf}{RLF}{Radio Link Failure}
\newacronym{rrc}{RRC}{Radio Resource Control}
\newacronym{rrm}{RRM}{Radio Resource Management}
\newacronym{rr}{RR}{Round Robin}
\newacronym{rs}{RS}{Remote Server}
\newacronym{rsrp}{RSRP}{Reference Signal Received Power}
\newacronym{rss}{RSS}{Received Signal Strength}
\newacronym{rtt}{RTT}{Round Trip Time}
\newacronym{rw}{RW}{Receive Window}
\newacronym{rx}{RX}{Receiver}
\newacronym{sa}{SA}{standalone}
\newacronym{sack}{SACK}{Selective Acknowledgment}
\newacronym{sap}{SAP}{Service Access Point}
\newacronym{sch}{SCH}{Secondary Cell Handover}
\newacronym{scoot}{SCOOT}{Split Cycle Offset Optimization Technique}
\newacronym{sdma}{SDMA}{Spatial Division Multiple Access}
\newacronym{sinr}{SINR}{Signal to Interference plus Noise Ratio}
\newacronym{sm}{SM}{Saturation Mode}
\newacronym{snr}{SNR}{Signal to Noise Ratio}
\newacronym{son}{SON}{Self-Organizing Network}
\newacronym{ss}{SS}{Synchronization Signal}
\newacronym{srs}{SRS}{Sounding Reference Signal}
\newacronym{sss}{SSS}{Secondary Synchronization Signal}
\newacronym{tb}{TB}{Transport Block}
\newacronym{tcp}{TCP}{Transmission Control Protocol}
\newacronym{tdd}{TDD}{Time Division Duplexing}
\newacronym{tdma}{TDMA}{Time Division Multiple Access}
\newacronym{tfl}{TfL}{Transport for London}
\newacronym{tm}{TM}{Transparent Mode}
\newacronym{trp}{TRP}{Transmitter Receiver Pair}
\newacronym{tti}{TTI}{Transmission Time Interval}
\newacronym{ttt}{TTT}{Time-to-Trigger}
\newacronym{tx}{TX}{Transmitter}
\newacronym{ue}{UE}{User Equipment}
\newacronym{ul}{UL}{Uplink}
\newacronym{uml}{UML}{Unified Modeling Language}
\newacronym{um}{UM}{Unacknowledged Mode}
\newacronym{utc}{UTC}{Urban Traffic Control}
\newacronym{vm}{VM}{Virtual Machine}
\newacronym{rsrq}{RSRQ}{Reference Signal Received Quality}
\newacronym{rssi}{RSSI}{Received Signal Strength Indicator}
\newacronym{crs}{CRS}{Cell Reference Signal}
\newacronym{comp}{CoMP}{Coordinated Multi-Point}
\newacronym{cran}{C-RAN}{Cloud \acrlong{ran}}
\newacronym{ca}{CA}{Carrier Aggregation}
\newacronym{cco}{CC}{Carrier Component}
\newacronym{nsa}{NSA}{Non Stand Alone}
\newacronym{embb}{eMBB}{Enhanced Mobility Broadband}
\newacronym{bsr}{BSR}{Buffer Status Report}
\newacronym{srb}{SRB}{Service Radio Bearer}
\newacronym{scm}{SCM}{Spatial Channel Model}
\newacronym{sctp}{SCTP}{Stream Control Transmission Protocol}
\newacronym{mptcp}{MPTCP}{Multi-path TCP}
\newacronym{ietf}{IETF}{Internet Engineering Task Force}
\newacronym{os}{OS}{Operating System}
\newacronym{tls}{TLS}{Transport Layer Security}
\newacronym{rfc}{RFC}{Request for Comments}
\newacronym{http}{HTTP}{HyperText Transfer Protocol}
\newacronym{nat}{NAT}{Network Address Translation}
\newacronym{api}{API}{Application Programming Interface}
\newacronym{rto}{RTO}{Retransmission Timeout}
\newacronym{psc}{PSC}{Public Safety Communication}
\newacronym{rpgm}{RPGM}{Reference Point Group Mobility}
\newacronym{ic}{IC}{Incident Command}
\newacronym{rsu}{RSU}{Road Side Unit}
\newacronym{uav}{UAV}{unmanned aerial vehicle}
\newacronym{usv}{USV}{Unmanned Surface Vehicle}
\newacronym{uas}{UAS}{Unmanned Aerial System}
\newacronym{iab}{IAB}{Integrated Access and Backhaul}
\newacronym{qoe}{QoE}{Quality of Experience}
\newacronym{ssim}{SSIM}{Structural Similarity Index}
\newacronym{psnr}{PSNR}{Peak Signal to Noise Ratio}
\newacronym{bs}{BS}{Base Station}
\newacronym{mu}{MU}{Multiple User}
\newacronym{ag}{AG}{Air-to-Ground}
\newacronym{af}{AF}{Array Factor}
\newacronym{ula}{ULA}{Uniform Linear Array}
\newacronym{upa}{UPA}{Uniform Planar Array}
\newacronym{lcs}{LCS}{Local Coordinate System}
\newacronym{psd}{PSD}{Power Spectral Density}
\newacronym{vq}{VQ}{vector quantization}
\newacronym{a2g}{A2G}{air-to-ground}
\newacronym{em}{EM}{electromagnetic}
\newacronym{vae}{VAE}{variational autoencoder}

\makeatletter

 \let\oldforeign@language\foreign@language
 \DeclareRobustCommand{\foreign@language}[1]{%
   \lowercase{\oldforeign@language{#1}}}

\usepackage[caption=false,font=footnotesize]{subfig}
\usepackage[font=small]{caption}%https://www.overleaf.com/project/5dea7e8d55668600017637bf

\usepackage{dblfloatfix}

\def\nbd{{\mathbf{d}}}

\def\nbu{{\mathbf{u}}}
\def\nbv{{\mathbf{v}}}

\def\nbx{{\mathbf{x}}}
\def\nby{{\mathbf{y}}}
\def\nbz{{\mathbf{z}}}
\def\nb0{{\mathbf{0}}}
\def\nb1{{\mathbf{1}}}

\makeatother

\IEEEoverridecommandlockouts

\begin{document}
\title{Generative Neural Network Channel Modeling  \\ for Millimeter-Wave UAV Communication}
%Millimeter Wave Channel Modeling \\ via Generative Neural Networks
% data-driven channel modeling for...
% 
%\author{
%\IEEEauthorblockN{\vspace{1cm} Blind review}
%
%\IEEEauthorblockN{William Xia$^{\dagger}$ \quad Sundeep Rangan$^{\dagger}$ 
%\quad Marco Mezzavilla$^{\dagger}$ \quad Angel Lozano$^{\flat}$ \\ \quad Giovanni Geraci$^{\flat}$ \quad Vasilii Semkin$^{\sharp}$ \quad Giuseppe Loianno$^{\dagger}$} \\
%\IEEEauthorblockA{$^{\dagger}$NYU Tandon School of Engineering, Brooklyn, NY, USA} \\
%\IEEEauthorblockA{$^{\sharp}$VTT Technical Research Centre of Finland Ltd, Finland} \\
%\IEEEauthorblockA{$^{\flat}$Univ. Pompeu Fabra, Barcelona, Spain} 

%\IEEEauthorblockN{William Xia, Sundeep Rangan, Marco Mezzavilla, Angel Lozano,\\Giovanni Geraci, Vasilii Semkin, and Giuseppe Loianno} \\
\author{William~Xia,~\IEEEmembership{Student~Member,~IEEE,}
        Sundeep~Rangan,~\IEEEmembership{Fellow,~IEEE,}
        Marco~Mezzavilla,~\IEEEmembership{Senior~Member,~IEEE,}
        Angel~Lozano,~\IEEEmembership{Fellow,~IEEE,}
        Giovanni~Geraci,~\IEEEmembership{Senior~Member,~IEEE,}
        Vasilii~Semkin,
        and~Giuseppe~Loianno,~\IEEEmembership{Member,~IEEE}% <-this % stops a space
\thanks{W. Xia, S. Rangan, M. Mezzavilla, and G. Loianno are with NYU Tandon School of Engineering, Brooklyn, USA. Their work is supported by NSF grants  1302336,  1564142,  1547332, and 1824434,  NIST, SRC, and the industrial affiliates of NYU WIRELESS.}
\thanks{A. Lozano and G. Geraci are with Univ. Pompeu Fabra, Barcelona. Their work is supported by ERC grant 694974, by MINECO's Project RTI2018-101040, by ICREA, and by the Junior Leader Fellowship Program from ``la Caixa" Banking Foundation.}
\thanks{V. Semkin is with VTT Technical Research Centre of Finland Ltd, Finland. His work is supported in part by the Academy of Finland.}
\thanks{Some of the material in this paper was presented at the 2020 IEEE GLOBECOM conference \cite{XiaRanMez2020}.}
%\thanks{W.~Xia, S.~Rangan, and M.~Mezzavilla were supported by NSF grants  1302336,  1564142,  1547332, and 1824434,  NIST, SRC, and the industrial affiliates of NYU WIRELESS. A.~Lozano and G.~Geraci were supported by ERC grant 694974, by MINECO's Project RTI2018-101040, by ICREA, and by the Junior Leader Fellowship Program from ``la Caixa" Foundation. The work of V.~Semkin is supported in part by the Academy of Finland.}
% \vspace{-8mm}
}

\maketitle

\begin{abstract}
The millimeter wave bands are being increasingly considered for wireless communication to unmanned aerial vehicles (UAVs).  
Critical to this undertaking are statistical channel models that describe the distribution of constituent parameters in scenarios of interest. 
This paper presents a general modeling methodology based on
data-training a generative neural network.
The proposed generative model has a two-stage structure that first
predicts the link state 
(line-of-sight, non-line-of-sight, or outage), and subsequently feeds this state into a conditional 
\gls{vae} that generates the path losses, delays, and angles of arrival and departure for all the propagation paths.
%Importantly, minimal prior assumptions are made, enabling the model to capture complex relationships within the data.
The methodology is demonstrated for \SI{28}{GHz} air-to-ground channels between UAVs and a cellular system in representative urban environments, with training datasets produced through ray tracing.
The demonstration extends to both standard base stations (installed at street level and downtilted) as well as dedicated base stations (mounted on rooftops and uptilted). 
The proposed approach is able to capture complex statistical relations in the data
and it significantly outperforms standard 3GPP models,
even after refitting the parameters of those models to the data.
\end{abstract}


\IEEEpeerreviewmaketitle

\section{Introduction}
\label{sec:intro}

Communication with \glspl{uav} is a subject of growing interest, and
the millimeter wave (mmWave) range is an inviting realm for this purpose because of the enormous bandwidth availability and the possibility of \gls{los} situations  \cite{GerGarAza2021,ZenGuvZha2020,SaaBenMoz2020,3GPP22125,3GPP22829,GerGarLin2019,FotQiaDin2019,MozSaaBen2018,GarGerLop2019,GerGarGal2018,LinWirEul2019,ZenLyuZha2019,AzaRosPol2019,NguAmoWig2018,SinBhaOzt2021,KanMezLoz2021}.
As with all communication systems, 
the design and evaluation of mmWave UAV networks hinges crucially on the availability of suitable channel models.

As current \gls{3gpp} channel models, which extend up to \SI{100}{GHz} for terrestrial users, are only calibrated for UAVs at sub-\SI{6}{GHz} frequencies \cite{3GPP36777}, there is a pressing need to extend the availability of channel models suitable for UAVs to the mmWave range.
For example, \cite{polesea2a} proposes a propagation model for UAV-to-UAV communication at \SI{60}{GHz} in LOS conditions, and with UAV altitudes ranging between 6 and \SI{15}{m}. 
Several other works have also attempted to model
various aggregate statistics of the channel model, such as 
the onmidirectional path loss or narrowband fading \cite{shakhatreh2021modeling,GarMohJai2020,KovMolSam2018,dabiri2019analytical,gapeyenko2018flexible}. 
%underlying measurements does not include NLOS links and hence does not characterize reflections and diffraction in the face of blockage.
%However, these channel models are not fully adequate for  many mmWave link and network simulations.
More generally, as mmWave systems rely on highly directional 
communication over wide bandwidths,
statistical descriptions of 
the full \emph{double directional} characteristics 
of the channel are required, meaning a description of the totality of path components (angles of arrival and departure, gains
and delays).
%Channel models have evolved over time and they have come to adopt, as their preferred form,
%with their preferred form having long been that of
%that of parametric statistical descriptions: distributions are provided for the angles of arrival and departure, the gains, and the delays of multipath components; 

Statistical channel models
%including industrial ones such as \cite{3GPP38901} typically provide methods to generate 
enable producing
random instances of the full set of channel parameters.
%by sampling these distributions, the channel response at any time/frequency/location can be realized.
%However, modeling the joint statistical distribution of these parameters (path angles, gains, and delays)  can be challenging.  
The joint statistical distribution of these parameters (path angles, gains, and delays) must first be distilled from a combination of physical considerations and field measurements \cite{rangan2014millimeter,Rappaport2014-mmwbook}, a process that
has become increasingly cumbersome as the systems being modelled have grown in complexity and heterogeneity (new frequency bands, broader bandwidths, massive antenna arrays, diverse deployments) \cite{wu2017general}.
In aerial settings, this complexity is further compounded by additional parameter dependencies on
the \gls{uav} altitudes, their 3D orientation, or the building heights, among others \cite{3GPP36777,3GPP38901,KhaGuvMat2019,AmoNguMog2017,9136077,SemKanHaa2021}. 
Altogether, the model parameters are bound to exhibit decidedly complex relationships that are  difficult to establish through analytical or physical considerations.  %This fact has 
%made the modeling process from challenging 
%and necessarily increasingly based on empirical data. 

% ANGEL'S ADDITIONAL MATERIAL: CHANNEL MODELS DO NOT INTEND TO PERFECTLY REPRODUCE REALITY (WOULDN'T BE POSSIBLE ANYWAY). RATHER, THEY INTEND TO CAPTURE ALL THE EFFECTS THAT ARE RELEVANT TO COMPARISONS OF COMPETING SYSTEMS, TECHNIQUES, IDEAS, ETC 

%Once it is accepted that the parameter distributions need to emanate from empirical data, with minimal room for other considerations, 
Modern data-driven machine-learning methods become an attractive recourse whenever 
physically based modeling is difficult.
Importantly, these methods entail minimal assumptions and can naturally capture intricate probabilistic relationships.
In such spirit, this paper considers data-driven methods to model mmWave air-to-ground channels.

Neural networks (NNs) have been advocated in \cite{stocker1993neural,chang1997environment,bai2018predicting,huang2018big,9122601} for indoor mmWave channel modeling, whereby, upon an input corresponding to some location, the NN outputs the model parameters for that location; in essence, the parameters are then a regression from the training dataset, much as in data-based signal power maps and in learning-based planning and prediction tools \cite{ostlin2010macrocell,dall2011channel,azpilicueta2014ray,kasparick2015kernel,ferreira2016improvement,romero2017learning,ma2017data,8647510}.
A strong aspect of all these works is their inherent site-specific nature, a virtue when it comes to optimizing specific deployments.
Alternatively, there is interest in models that can produce channel parameters broadly representative of some general environment, say an urban microcellular system.
%The other side of the coin is that the models thus produced lack generality, which is a desirable attribute for the purposes of research, development, algorithm comparison, and even standardization. 

Generative NNs, which have proven enormously successful with images and text \cite{radford2015unsupervised,goodfellow2014generative,doersch2016tutorial}, 
offer a natural approach to data-driven channel modeling that can broadly represent complex settings, and
some early works have successfully trialed generative adversarial networks (GANs) for simple wireless channels \cite{8663987,8685573,8985539}.
The present paper propounds a different generative NN structure, powerful and  widely applicable, for air-to-ground channel modeling.
For data provisioning, we rely on the ray tracing tool \cite{Remcom},
which has developed substantially for mmWave communication
\cite{Alkhateeb2019,degli2014ray,HouKanJam2014,YunIsk2015,khawaja2017uav,fuschini2017analysis} and can supply datasets of the size required to train large NNs.

Ray tracing requires a detailed blueprint of the environment, including the size, shape, and location of all obstacles, along with their electromagnetic properties. As it employs high-frequency approximations, ray tracing exhibits some inaccuracies,
but is perfectly adequate for our purpose here, which is to validate the proposed methodology.
%As it employs high-frequency approximations, ray tracing is not as accurate as full wave electromagnetic solutions, but it is far less complex and it can provide satisfactory predictions for path loss, angle of arrival/departure, delay profile, and wideband results.
We hasten to emphasize that,
%the purpose of the ray-tracing data is to validate the proposed structure.
%but that
ultimately, the model is meant to be driven by field data, gathered either through targeted measurement campaigns or directly supplied by users of the service.

The highlights of this work are as follows:
\begin{itemize}[leftmargin=*]
    \item \emph{Double-directional wideband characterization}.
    As chief point, we demonstrate that the proposed
    method can capture the directional characteristics of the channel at both transmitter and receiver along with its wideband nature,
    meaning the angular, gain, and delay information for all the paths on each link. 
    This description is compatible with 3GPP evaluation 
    methodologies \cite{3GPP38901,3GPP36777} and can provide the full
    wideband MIMO response given specific antenna configurations at transmitter
    and receiver. No prior assumptions are made
    regarding the dependencies among parameters, and
    the model is able to capture relationships that are nuanced and interesting.
    
    \item \emph{Novel NN structure}.
    The generative model features a novel two-stage
    structure where a first NN determines
    whether the link is in a state of LOS, non-line-of-sight (NLOS), or outage, while a second stage employs a conditional \gls{vae} to 
    generate the path parameters given that state.  
    Importantly, several pre-processing steps are introduced to map the path parameters
    to a format compatible with NN outputs.

\item \emph{Application to UAV mmWave settings}. The methodology is demonstrated by characterizing
\SI{28}{GHz} channels connecting UAVs with two distinct classes of ground base stations.
%standard (installed at street level, downtilted) and dedicated (installed on rooftops, uptilted) base stations. 

\item \emph{Intra- and inter-environment 
generalization}.
The model is separately trained on data from various environments, namely sections of Tokyo, Beijing, London, Moscow, and Boston. Then, these models are tested on new points from the respective datasets as well as on points from the other datasets.
%two environments, namely (\emph{i}) an environment featuring high-rise buildings as well as wide open areas, embodied by sections of Tokyo and Beijing, and (\emph{ii}) an environment dominated by low-rise buildings, represented by sections of London and Moscow.
This allows testing the ability of the model trained in one environment to describe the behavior in new locations within that environment (intra-environment generalization) and in locations in other environments (inter-environment generalization).
%We train models on data from two pairs of cities:  (i) Tokyo-Beijing representing the two cities with high buildings and (ii) London-Moscow representing two cities with lower buildings.
%We use these groups to test the ability of models trained in one environment to generalize to new locations within the same environment (intra-environment generalization) and new locations in  different environments (inter-environments).
%Interestingly, the link state distribution is seen to vary substantially across environments, while the path parameters conditioned on that state generalize well.

%we see that, among these two groups of cities, the link state models can vary between models, while the path model condition on the link state generalizes well.

\item \emph{Benchmarking against 3GPP models}.
The proposed generative model is benchmarked against the existing \gls{3gpp} channel model, recalibrated to fit the mmWave data used to train our generative model.
The generative model proves superior, highlighting the advantage of techniques that make minimal prior structural assumptions.

\item \emph{Publicly available model}.
The developed model is publicly available \cite{mmw-github} and can be readily incorporated to any simulator of mmWave UAV communication.
And, beyond this use case, the underlying modeling framework may be enticing for other emerging communication scenarios such as terahertz systems, and even as an alternative to traditional models in other contexts.

\end{itemize}

The paper is organized as follows. Section \ref{sec:gen} frames the problem, Section \ref{sec:model} sets forth the proposed generative approach, and Section \ref{sec:remcom} describes the data procurement process. Then, Section \ref{sec:results} presents a battery of results that illustrate how the trained model successfully predicts the channel's behavior in unseen locations. Finally, Section \ref{subsec:compare_3gpp} contrasts the predictive power of the proposed model against that of the refitted 3GPP model, and Section \ref{sec:conclusion} concludes the paper.

%\vspace{3mm}
\section{Problem Formulation}
\label{sec:gen}

We consider the modeling of channels linking a transmitter with a
receiver. The UAV is taken to be the transmitter and the base station---gNB in 3GPP terminology
\cite{3GPP36777}---the receiver, yet, owing to reciprocity,
the roles of transmitter and receiver are interchangeable.  
Each link is described by the collection of parameters \cite{heath2018foundations}
\begin{equation} \label{eq:xvec}
    \nbx = \Big \{ \left( L_k,\phi^{\rm rx}_k,\theta^{\rm rx}_k,
    \phi^{\rm tx}_k,\theta^{\rm tx}_k,
    \tau_k \right), ~
    k=1,\ldots,K \Big \},
\end{equation}
where $K$ is the number of paths
whereas $L_k$ is the loss of path $k$,
$(\phi^{\rm rx}_k,\theta^{\rm rx}_k)$
are its azimuth and elevation angles of arrival,
$(\phi^{\rm tx}_k,\theta^{\rm tx}_k)$
are its azimuth and elevation angles of departure, and 
$\tau_k$ is its absolute propagation delay.
Unlike 3GPP spatial cluster models
(e.g., \cite{3GPP38901}), we do not consider
angular or delay dispersion within each path. 
This is not a limitation of the model, but only a consequence
of the tool that produces training datasets
%(see Sec.~\ref{sec:remcom})
with discrete paths.
If angular or delay spread information were available, these aspects could be incorporated.

For the sake of specificity, the number of paths is fixed at $K = 20$, with $L_k=L_{\rm max}$
for paths that are not actually present; we set $L_{\rm max}=$\, \SI{200}{dB},
which is compatible with the maximum path loss
detectable by the ray tracer.
With these settings, 
the data vector $\nbx$ in \eqref{eq:xvec}
contains $6 K=120$ parameters per link.
Let
\begin{equation} \label{eq:uvec}
    \nbu = [\nbd, c]
\end{equation}
denote the \emph{link condition},
with $\nbd = [d_{\text{x}},d_{\text{y}},d_{\text{z}}]$ the vector connecting the UAV with the gNB
and with $c$ indicating the type of gNB.
For the air-to-ground modeling problem, we consider 
two types of gNBs:
\begin{itemize}
    \item \emph{Standard gNBs}, installed at street level and downtilted to serve terrestrial users, but potentially 
    usable for UAV connectivity; and
    \item \emph{Dedicated gNBs}, mounted on rooftops and uptilted,
    intended specifically for UAVs.
\end{itemize}
One could also consider other aspects, such as the gNB height,
within $c$; our methodology is general.

The goal is to capture the conditional distribution 
$p(\nbx|\nbu)$, %over some ensemble of possible links.
that is, to model the distribution
of the paths in a link as a function of that link's
condition in some environment.  
As anticipated, we consider
a generative scheme in which %we model $\nbx$ as 
\begin{equation} \label{eq:xgenuz}
    \nbx = g(\nbu, \nbz),
\end{equation}
where $\nbz $ is a random vector, termed the \emph{latent vector}, with some fixed prior distribution
$p(\nbz)$, while $g(\nbu, \nbz)$
is the \emph{generating function}, to be trained from data.

Once trained, 
generative models are conveniently applicable in 
simulations: the locations of UAVs and gNBs
are determined, either deterministically or stochastically according to some deployment 
strategy, providing the condition
vector $\nbu$ for each link.  Random vectors $\nbz$ can then be produced for each link from
the prior $p(\nbz)$ and, from $\nbu$ and $\nbz$, 
the path parameters $\nbx$ follow as per (\ref{eq:xgenuz}).  
These parameters can be generated for both
intended and interfering links and, in conjunction with the antenna patterns, array configuration, and beam tracking methods, allow computing quantities of interest such as signal-to-noise ratios (SNRs), signal-to-interference-plus-noise ratios (SINRs), or bit rates.

Small-scale dynamics
%large scale parameters can be assumed stationary,
can also be modeled under the premise of local stationarity. %from these parameters.
Specifically, given any local motion with some
velocity, Doppler shifts can be computed and applied to each path to derive the
time-varying wideband frequency response~\cite{heath2018foundations}.  
However, statistical modeling of
large-scale dynamics such as blockage \cite{maccartney2017rapid,slezak2018empirical}
and spatial consistency \cite{3GPP38901} remain an interesting avenue
of future research.

%\vspace{3mm}
\section{Proposed Generative Model}
\label{sec:model}
\subsection{Overview}
\begin{figure*}[t]
\footnotesize
\centering
\begin{tikzpicture}[scale=0.85,every text node part/.style={align=center}, every node/.append style={transform shape}]

    \def\layersep{2.5cm}

    \tikzstyle{every pin edge}=[<-,shorten <=1pt]
    \tikzstyle{neuron}=[circle,fill=black!25,minimum size=17pt,inner sep=0pt]
    \tikzstyle{input neuron}=[neuron, fill=green!50];
    \tikzstyle{output neuron}=[neuron, fill=red!50];
    \tikzstyle{hidden neuron}=[neuron, fill=blue!50];
    \tikzstyle{annot} = [text width=4em, text centered]

    % Draw the input layer nodes
%    \foreach \name / \y in {1,...,5}
    % This is the same as writing \foreach \name / \y in {1/1,2/2,3/3,4/4}
%        \node[input neuron, pin=left:Input \#\y] (I-\name) at (0,-\y) {};
    \node[input neuron, pin=left:Condition Vector $\nbu$] (I-11) at (0,-1) {};
%    \node[input neuron, pin=left:Horizontal Distance] (I-11) at (0,-1) {};
%    \node[input neuron, pin=left:Vertical Distance] (I-12) at (0,-2) {};
%    \node[input neuron, pin=left:Base Station Type] (I-13) at (0,-3) {};
            
    % Draw the hidden layer nodes
    \foreach \name / \y in {1,...,2}
        \path[yshift=0.5cm]
            node[hidden neuron,xshift=-0.5cm] (H1-\name) at (\layersep,-\y cm) {};            
    
    \draw node[label,below of=H1-2] (H1-3) {\Large $\vdots$};
    
    % Draw the output layer node
    \node[output neuron,right of=H1-2,xshift=1cm,yshift=0.5cm] (O) {};
    
    \node [below of=O]  (zs) {$z_{\rm state}$};
    \draw [->] (zs) -- (O.south);
    
    % Connect every node in the input layer with every node in the
    % hidden layer.
    \foreach \source in {1,...,1}
        \foreach \dest in {1,...,3}
            \path (I-1\source) edge (H1-\dest);

    % Connect every node in the hidden layer with the output layer
    \foreach \source in {1,...,3}
        \path (H1-\source) edge (O);

    % Annotate the layers
    % Link Conditions
%    \node [draw,fill=green!40, above left of=I-11, node distance=1.5cm,xshift=-1cm] 
%    (input) {Link Conditions};
    
%    \node[annot,above of=H1-1,node distance=1cm] (h) {Hidden layer};
%    \node[annot,left of=h,node distance=2cm,xshift=0.5cm] (i) {Input layer};
%    \node[annot,right of=h, node distance=2cm] {Output layer};
    
    \node[input neuron,right of=O,xshift=3cm] (I-21) {};
    \path (O) edge (I-21);
    
    \node[annot,above left of=I-21, xshift=-1.25cm] {Link State $s$};
    
    \node[input neuron, below of=I-21, pin=left:Condition Vector $\nbu$] (I-22) {};
    \node[input neuron, below of=I-22, pin=left:Latent Variable $\nbz_{\rm NLOS}$] (I-23) {};
%    \node[input neuron, below of=I-21, pin=left:Horizontal Distance] (I-22) {};
%    \node[input neuron, below of=I-22, pin=left:Vertical Distance] (I-23) {};
%    \node[input neuron, below of=I-23, pin=left:Base Station Type] (I-24) {};
%    \node[input neuron, below of=I-24, pin=left:Random $z$] (I-25) {};
    
    % Draw the hidden layer nodes
    \node[hidden neuron,right of=I-21,yshift=0.5cm,xshift=1cm] (H2-1) {};    
    \node[hidden neuron,right of=I-22,yshift=0.5cm,xshift=1cm] (H2-2) {};    
    \node[hidden neuron,right of=I-23,yshift=0.5cm,xshift=1cm] (H2-3) {};    
    \node[hidden neuron,below of=H2-3] (H2-4) {};    
%    \node[hidden neuron,below of=H2-4] (H2-5) {};    
%    \node[hidden neuron,below of=H2-5] (H2-6) {};    
    \draw node[label,below of=H2-4] (H2-5) {\Large $\vdots$};   
    
    \foreach \source in {1,...,3}
        \foreach \dest in {1,...,5}
            \path (I-2\source) edge (H2-\dest);
    
    \node[output neuron,pin={[pin edge={->}]right:Path Parameters $\nbx$},right of=H2-1,yshift=-0.5cm,xshift=1cm] (O1) {};    
%    \node[output neuron,pin={[pin edge={->}]right:Path gains},below of=O1] (O2) {};    
%    \node[output neuron,pin={[pin edge={->}]right:Path angles},below of=O2] (O3) {};    
%    \node[output neuron,pin={[pin edge={->}]right:Path delays},below of=O3] (O4) {};    

    \foreach \source in {1,...,5}
        \foreach \dest in {1,...,1}
            \path (H2-\source) edge (O\dest);    
            
    \node [below of=O1]  (zo) {$\nbz_{\rm out}$};
    \draw [->] (zo) -- (O1.south);
    
    % Label networks
    \draw [decorate,decoration={brace,amplitude=10pt,raise=1cm}]
        (I-11.west) -- (O.east) node (link_network) {};
    \draw node[label,above of=link_network,xshift=-2.25cm,yshift=1cm] () 
        {Link-State Prediction \\ Stage};
        
    \draw [decorate,decoration={brace,amplitude=10pt,raise=1cm}]
        (I-21.west) -- (O1.east) node (vae_network) {};
    \draw node[label,above of=vae_network,xshift=-2.25cm,yshift=1cm] () 
        {Path Generation \\ Stage};    

    % Output label
%    \node [draw,fill=green!40, above right of=O1, node distance=1.5cm,xshift=1cm] 
%    (input) {Link Data Vector};
    
\end{tikzpicture}
\caption{Overall architecture for the two-stage generative model, which accepts a link condition vector $\nbu$ and a
latent vector 
$\nbz=[z_{\mathrm{state}},\nbz_{\mathrm{NLOS}},\nbz_{\mathrm{out}}]$ to generate random 
path parameters $\nbx$. (For the sake of clarity, various transformations, described in the text, are omitted from this diagram.)}
\label{fig:gen_model}
\vspace{-2mm}
\end{figure*}
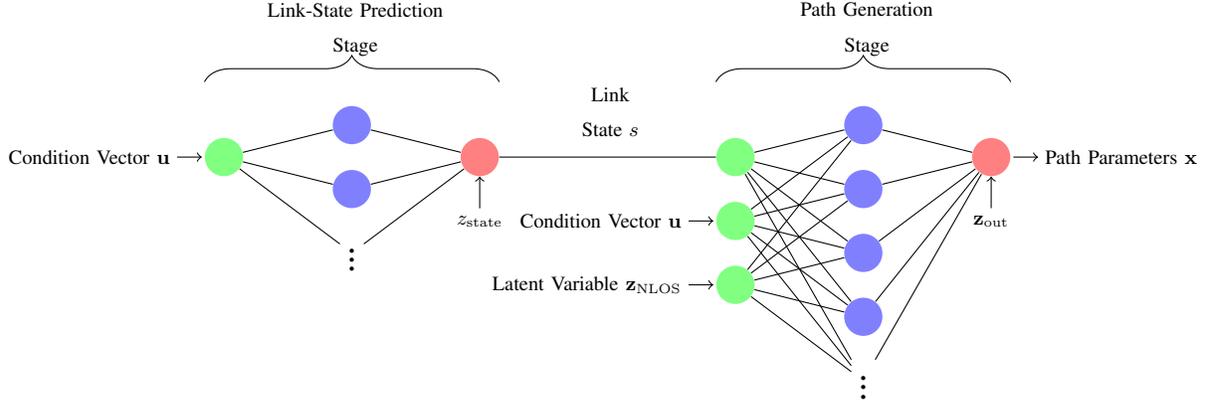

%As described above, the goal is to fit a generative model for the path vector $\nbx$ in \eqref{eq:xvec} of the form $\nbx = g(\nbu,\nbz)$ where $\nbu$ is the link condition vector \eqref{eq:uvec} and $\nbz$ is some random vector.
The propounded generative model, sketched in Fig.~\ref{fig:gen_model},
constructs the generative function 
as two cascaded stages, namely a link-state prediction stage followed by a path generation stage. The latent vector
$\nbz$ subsumes three components,
\begin{equation} \label{eq:zpart}
    \nbz = [z_{\mathrm{state}}, \nbz_{\mathrm{NLOS}}, \nbz_{\mathrm{out}}].
\end{equation}
The link-state predictor accepts the 
condition vector $\nbu$ and a random
variable $z_{\mathrm{state}}$, from which it determines the link state $s$. From $s$ and
the two other latent components,
$\nbz_{\mathrm{NLOS}}$ and $\nbz_{\mathrm{out}}$, the path generation stage then
produces the final path parameters $\nbx$.
We next describe the details of this whole architecture.

\subsection{Link-State Predictor}

As recognized by 3GPP models such as \cite{3GPP38901},
it is crucial to first determine the existence or lack of the LOS path.
To this end, the link-state predictor accepts the condition $\nbu$ defined
in \eqref{eq:uvec} and produces probabilities for the link being in one of three states \cite{akdeniz2014millimeter}:
\vspace{2mm}
\begin{itemize}%[leftmargin=*]
\item {\tt LOS}: The LOS path is present, 
possibly in addition to NLOS paths;
\item {\tt NLOS}: The LOS path is blocked, but at least one NLOS path is active;
\item {\tt NoLink}: No propagation paths (either LOS or NLOS) exist for this link.
\end{itemize}
In the sequel, $s \in  \{ \tt{LOS},\tt{NLOS},\tt{NoLink} \}$
denotes the predicted link state while the generative model mapping $\nbu$ to $s$ is represented
by
\begin{equation} \label{eq:gstate}
    s = g_{\mathrm{state}}(\nbu, z_{\mathrm{state}}).
\end{equation}
Such mapping entails three steps, expounded next.

%\emph{condition vector transformation},
%\emph{neural network}, and 
%\emph{sampling}.
%The details of the steps are as follows.

\medskip
\noindent
\subsubsection{Condition Vector Transformation} 
The vector $\nbu$
is transformed into a new vector
\begin{equation} \label{eq:utrans}
    \Big[ \mathbf{c}_{\mathrm{one}}, d_{\rm 3D} \mathbbm{1}_{\{c=1\}}, 
    d_{z} \mathbbm{1}_{\{c=1\}}, \cdots, 
    d_{\rm 3D} \mathbbm{1}_{\{c=C\}}, 
    d_{z} \mathbbm{1}_{\{c=C\}} \Big]
\end{equation}
where $\mathbf{c}_{\mathrm{one}}$ is a one-hot
coded version of the gNB type $c$ while
$d_z$ is the vertical distance, 
\begin{equation}
d_{\rm 3D}=\sqrt{d_x^2+d_y^2+d_z^2}
\end{equation}
is the 3D
distance, $C$ is the number of possible gNB types,
and $\mathbbm{1}_{\{c=i\}}$ is the indicator
function for the event $c=i$.
As $c$ can take $C$ possible values,
we can one-hot code $\mathbf{c}_{\mathrm{one}}$ with
$C-1$ dimensions.  Hence, the transformed vector in \eqref{eq:utrans}
has dimension $C-1+2C=3C-1$.
With $C=2$ (standard and dedicated),
the transformed vector 
in \eqref{eq:utrans} has $3C-1=5$ components.
The motivation for the 
transformation in \eqref{eq:utrans}
is to enable a different
behavior of the first layer of the NN for different types of gNB.

The transformed vector in \eqref{eq:utrans}
is passed through a min-max scaler that
maps its components to values between 0 and 1;
the limits on this min-max scaler are learned 
during training.  The resulting
transformed and scaled value is denoted by $\nbv_{\mathrm{state}}$.

\medskip
\noindent
\subsubsection{NN}
A fully connected NN, configured as per Table~\ref{tab:arch_param}, generates the link-state probabilities.
The input to this NN is $\nbv_{\mathrm{state}}$ while its output is a three-way softmax corresponding to the three states.  

\medskip
\noindent
\subsubsection{Sampling}
In the final step, a uniform
random variable $z_{\mathrm{state}} \in [0,1]$ samples the link state $s$ based 
on the probability outputs from the NN.
%takes the three probabilities and a uniform random variable $z_{\mathrm{state}} \in [0,1]$ to randomly sample the link state $s$ based  on the probabilty outputs from the NN.

\begin{table}[h]
    \vspace{1mm}
    \caption{Generative model configuration}
    \label{tab:arch_param}
    \centering
    \begin{tabular}{l|c|c|c|}
    \cline{2-4}
    %\multirow{3}{*}{} & 
    %\multicolumn{3}{c|}{}  \\ \cline{2-4}
    %{} &
    %\shortstack{Link state \\
    %prediction} & \shortstack{Path VAE\\ encoder} &
    %\shortstack{Path VAE\\ decoder} \\ \hline
    {} &
    \shortstack{Link state} & \shortstack{Path VAE} &
    \shortstack{Path VAE} \\
    {} &
    \shortstack{prediction} & \shortstack{encoder} &
    \shortstack{decoder} \\ \hline
    \multicolumn{1}{|l|}{Number of inputs} &
    5 & 5 + 120 & 5 + 20 \\ \hline
    \multicolumn{1}{|l|}{Hidden units} &
    $[25,10]$ & $[200,80]$ & $[80,200]$ \\ \hline
    \multicolumn{1}{|l|}{Number of outputs} &
    3 & $20+20$ & $120+120$ \\ \hline
    \multicolumn{1}{|l|}{Optimizer} &
    Adam & \multicolumn{2}{c|}{Adam} \\ \hline
    \multicolumn{1}{|l|}{Learning rate} &
    $10^{-3}$ & \multicolumn{2}{c|}{$10^{-4}$}
    \\ \hline
    \multicolumn{1}{|l|}{Epochs} &
    50 & \multicolumn{2}{c|}{10000} \\ \hline
    \multicolumn{1}{|l|}{Batch size} &
    100 & \multicolumn{2}{c|}{100} \\ \hline
    \multicolumn{1}{|l|}{Number of NN parameters} &
    1653 & 44520 & 40720 \\ \hline
    \end{tabular}
\noindent
\end{table}

\subsection{Path Generation Stage}

The second stage generates
the parameters $\nbx$ in \eqref{eq:xvec} 
given $\nbu$ and $s$.  
This also entails various steps, described next.

\subsubsection{Condition Vector Transformation} 
Again, we begin by transforming  $\nbu$ and $s$, in this case into
\begin{equation} \label{eq:utrans_path}
    \Big[ \mathbf{c}_{\mathrm{one}}, d_{\rm 3D}, 
    10\log_{10}(d_{\rm 3D}), d_z,
    s \Big],
\end{equation}
where $\mathbf{c}_{\mathrm{one}}$, $d_{\rm 3D}$ 
and $d_z$ are as in \eqref{eq:utrans}.
For this condition vector, we found that including both
$d_{\rm 3D}$ and $\log_{10}(d_{\rm 3d})$ enabled better
modeling with a smaller NN.
%That is, $c_{\mathrm{one}}$ is the one-hot coded version of the cell-type, $d_{3D}$ is the 3D distance and $d_z$ is the vertical distance.
This five-dimensional
vector is then passed through a min-max
scaler to produce a five-dimensional vector
with values between 0 and 1.
We denote this transformed
vector by $\nbv_{\mathrm{path}}$.

\subsubsection{NLOS VAE}
The next, and most intricate step, is to
generate the parameters for the NLOS paths
within $\nbx$.
As explained below, 
these NLOS paths are represented in a transformed version denoted
by $\nby_{\mathrm{NLOS}}$.
%The details and the motivation of this transformation will be explained below.
For now, we recall that there are up to $K=20$ NLOS paths with 6 parameters per path,
meaning that $\nby_{\mathrm{NLOS}}$ is of dimension $6K=120$.

We want to generate $\nby_{\mathrm{NLOS}}$ 
from $\nbv_{\mathrm{path}}$ and from some randomness.
This mapping should be trained such that the conditional
distribution of $\nby_{\mathrm{NLOS}}$
given $\nbv_{\mathrm{path}}$ matches the 
distribution in the training dataset.
There are a large number of methods for training 
generative models, the two most common being variants of 
GANs \cite{radford2015unsupervised,goodfellow2014generative}
or VAEs \cite{doersch2016tutorial}.  
We found the most success with a VAE, as it avoids the minimax optimization required by a GAN.

We apply a standard VAE architecture \cite{doersch2016tutorial} that has itself two stages:
the first stage accepts as inputs a random vector $\nbz_{\mathrm{NLOS}}$ along with $\nbv_{\mathrm{path}}$
and it outputs means and variances for the NLOS components, namely
\begin{equation}
  [\boldsymbol{\mu}_y, \boldsymbol{\sigma}^2_y] = 
  g_{\rm NLOS}(\nbv_{\mathrm{path}},\nbz_{\rm NLOS}).
\label{eq:gnlos}
\end{equation}
The vectors $\boldsymbol{\mu}_y$ and $\boldsymbol{\sigma}^2_y$ share the
dimensions of the sought $\nby_{\mathrm{NLOS}}$, hence they combine into 120+120 output values.
The entries of $\nbz_{\rm NLOS}$ are i.i.d.
Gaussian with mean zero and unit variance.
In VAE terminology, the dimension of 
$\nbz_{\rm NLOS}$ is termed the \emph{latent
dimension}, with  higher such dimensions enabling
better fitting to the data but requiring larger training datasets.
In the remainder, the latent dimension is kept at 20.

The sought $\nby_{\mathrm{NLOS}}$ is
sampled from the means and variances,
\begin{equation}
  \nby_{\rm NLOS} = \boldsymbol{\mu}_y + \boldsymbol{\sigma}_y \odot 
    \nbz_{\mathrm{out}},
    \label{eq:gout}
\end{equation}
where $\nbz_{\mathrm{out}}$ has 120 zero-mean
unit-variance i.i.d. Gaussian entries and $\odot$ indicates entry-wise
multiplication.

In the VAE paradigm, the generator in 
\eqref{eq:gnlos} 
is termed the \emph{decoder}. The VAE
also requires training a so-called \emph{encoder} that maps
data samples $\nby_{\rm NLOS}$ and $\nbv_{\rm path}$ back to the latent vector $\nbz_{\rm NLOS}$.  This encoder attempts to approximate
sampling from the posterior density of  $\nbz_{\rm NLOS}$ given
$\nby_{\rm NLOS}$ and $\nbv_{\rm path}$.
The encoder and decoder are then jointly optimized to maximize
an approximation of the log-likelihood called the evidence lower bound (ELBO);
see \cite{doersch2016tutorial} for details.  

Similar to standard VAE architectures
\cite{doersch2016tutorial}, we approximate the
posterior density of $\nbz_{\rm NLOS}$ given
$\nby_{\rm NLOS}$ and $\nbv_{\rm path}$ by a Gaussian
with independent components.
Hence, 
the encoder takes as inputs $\nby_{\rm NLOS}$ and $\nbv_{\rm path}$, and 
output a vector of mean and a vector of variances for the
latent variables $\nbz_{\rm NLOS}$. 
Under this assumption,
the encoder can be represented as a function
\begin{equation}
  [\boldsymbol{\mu}_z, \boldsymbol{\sigma}^2_z] = 
  h_{\rm NLOS}(\nbv_{\mathrm{path}},\nby_{\rm NLOS}),
\label{eq:hnlos}
\end{equation}
that takes as inputs $\nby_{\rm NLOS}$ and $\nbv_{\rm path}$ and outputs a vectors
$\boldsymbol{\mu}_z$ and
$\boldsymbol{\sigma}^2_z$ representing
the mean and variance of $\nbz_{\rm NLOS}$
given $\nby_{\rm NLOS}$ and $\nbv_{\rm path}$.
The vectors $\boldsymbol{\mu}_z$ and
$\boldsymbol{\sigma}^2_z$ will have the same dimension as the latent vector $\nbz_{\rm NLOS}$.  Given the outputs of the encoder, we can then sample from 
the approximate posterior density by
\begin{equation}
  \nbz_{\rm NLOS} = \boldsymbol{\mu}_z + \boldsymbol{\sigma}_z \odot 
    \boldsymbol{\epsilon},
    \label{eq:zsamp}
\end{equation}
where, again, $\odot$ represents elementwise
multiplication and $\boldsymbol{\epsilon}$
is i.i.d.\ zero-mean unit-variance Gaussian
noise.

In our case, the encoder and decoder
are embodied by fully connected NNs configured as per Table~\ref{tab:arch_param}.
Since the latent vector $\nbz_{\rm NLOS}$ is realized as a 20-dimensional Gaussian vector,
the decoder accepts this 20-dimensional Gaussian vector plus the
five-dimensional vector $\nbv_{\rm path}$
%transformed vector in (\ref{eq:utrans_path}), properly scaled,
and yields the 120+120 means and variances.
Conversely, the encoder is fed $\nbv_{\rm path}$
%the transformed vector in (\ref{eq:utrans_path})
and a 120-dimensional data input and produces means and variances 
for the 20-dimensional latent vector. 

\subsubsection{NLOS Transformation}
As advanced, the generated vector $\nby_{\rm NLOS}$ is a transformed
version of the path parameters, the reason being that those actual parameters
are heterogeneous: they include path losses, angles, and delays. To put them on an equal footing, $\nbx_{\rm NLOS}$ maps onto $\nby_{\rm NLOS}$ as follows:
\begin{itemize}
    \item The path losses are converted to dB-scale path gains and the minimum such value in the dataset is subtracted out. The resulting excess path gains are then run through a min-max scaler to lie between $0$ and $1$; a value of zero corresponds to the maximum path loss ($L_{\rm max}$) and hence to absence of this path altogether. 
        \item The angles are rotated relative to the LOS direction, and then scaled such that 180$^\circ$ corresponds to a unit value.
    \item The LOS delay is subtracted from the rest of delays, and the resulting excess delays are again scaled to be between $0$ and $1$.
\end{itemize}
The above transformations ensure that all values
are in a similar range and referenced to the LOS path.
The min-max scalers for the path losses and 
delays are fit to the training data, and we note that
the mapping of angles and delays relative
to the LOS path can take place even if such LOS path does not exist (because of blockage).

Once $\nby_{\mathrm{NLOS}}$ has been generated, the transformation must be undone to obtain the NLOS path parameters, $\nbx_{\mathrm{NLOS}}$.

\subsubsection{Addition of the LOS Path}
For the LOS path, when it exists, the delay and angles of departure and arrival can be computed from sheer geometry while its loss can be computed 
from Friis' law \cite{heath2018foundations}.
The final step is the addition, when it exists, of such LOS path to $\nbx_{\mathrm{NLOS}}$, which renders the full collection of path parameters, $\nbx$.
\section{Ray Tracing Data at 28 GHz}
\label{sec:remcom}

Experimental data on UAV channels is limited, 
particularly in the mmWave bands
\cite{KhaGuvMat2019,KhuCheZha2018,AmoNguMog2017,AmoMogSor2017,AmoNguWig2018}.
In this work, we employ a powerful ray tracing package, Wireless InSite by Remcom \cite{Remcom}, also used in \cite{khawaja2017uav,Alkhateeb2019}.

\begin{figure*}
\vspace{0.02\columnwidth}
\centering
\subfloat[Tokyo, Japan]{\includegraphics[width=0.4\columnwidth]{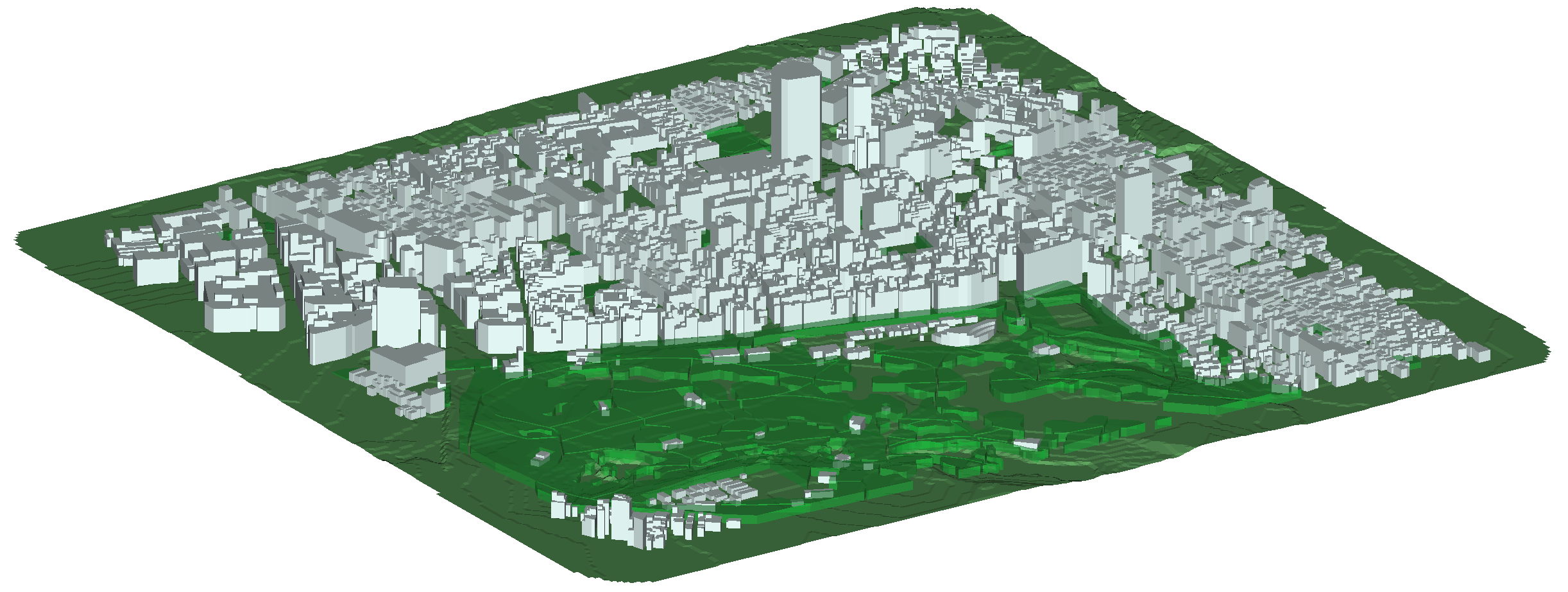}}\hspace{0.15\columnwidth}
\subfloat[Beijing, China]{\includegraphics[width=0.4\columnwidth]{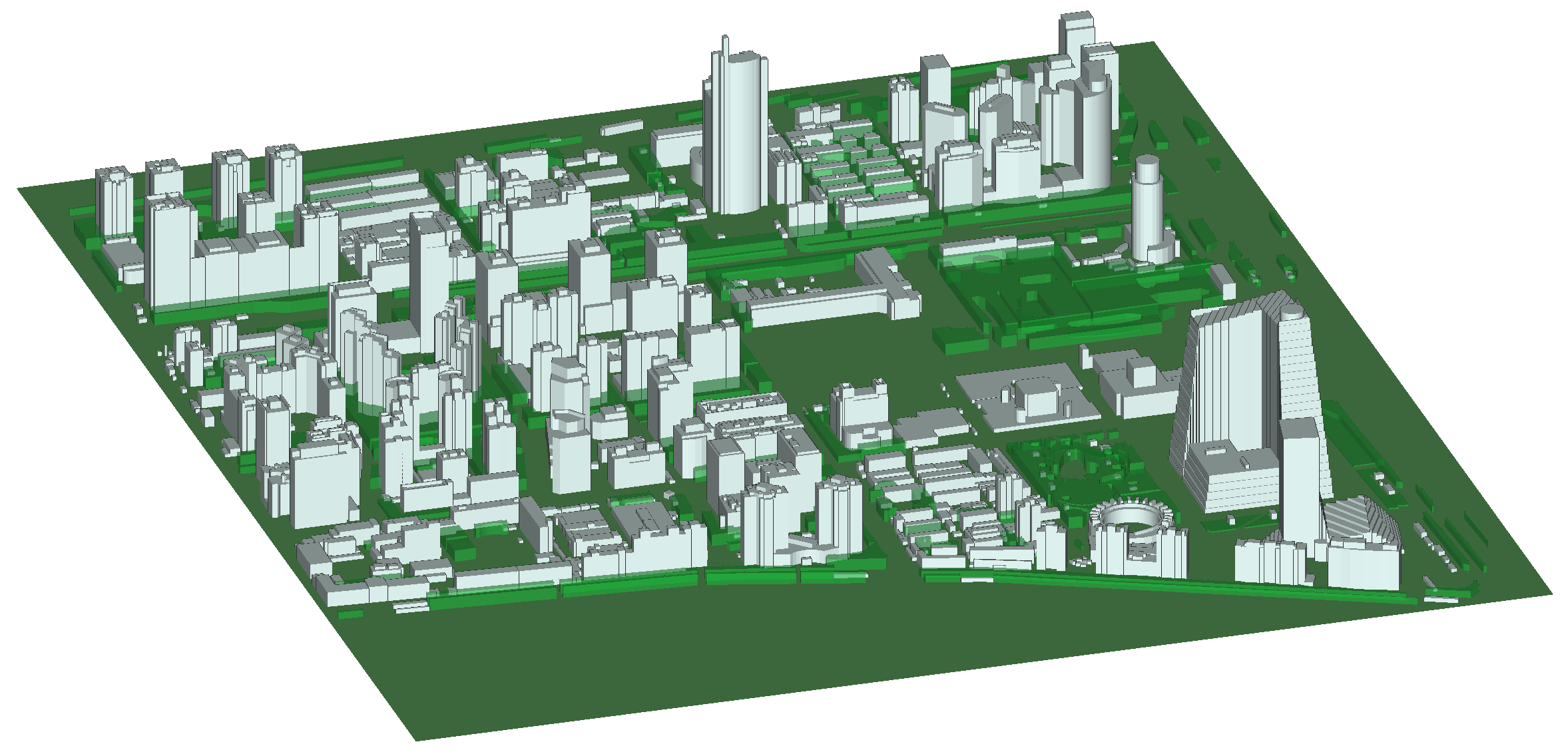}}\\\vspace{0.05\columnwidth}
\subfloat[London, UK]{\includegraphics[width=0.4\columnwidth]{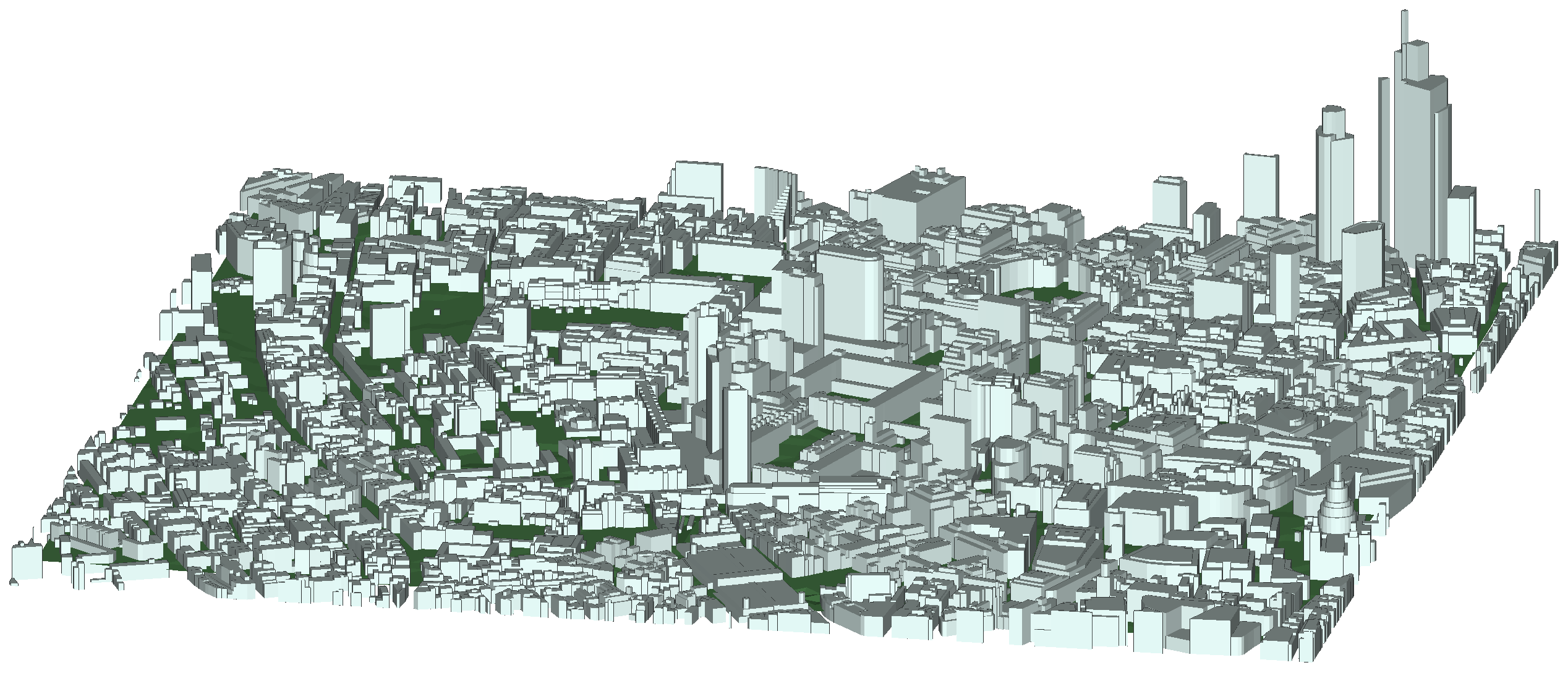}}\hspace{0.15\columnwidth}
\subfloat[Moscow, Russia]{\includegraphics[width=0.4\columnwidth]{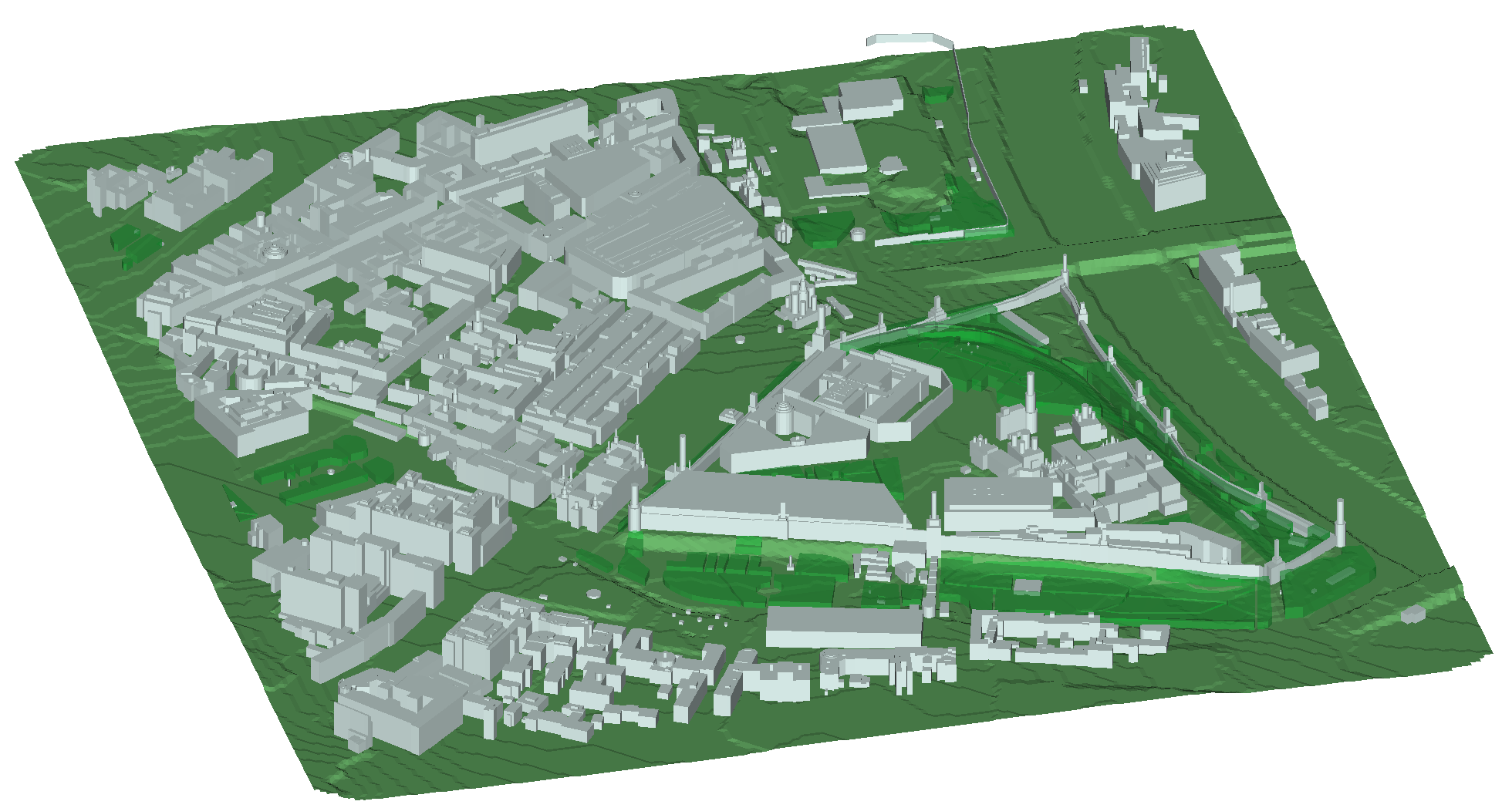}}\hspace{0.15\columnwidth}
\subfloat[Boston, USA]{\includegraphics[width=0.4\columnwidth]{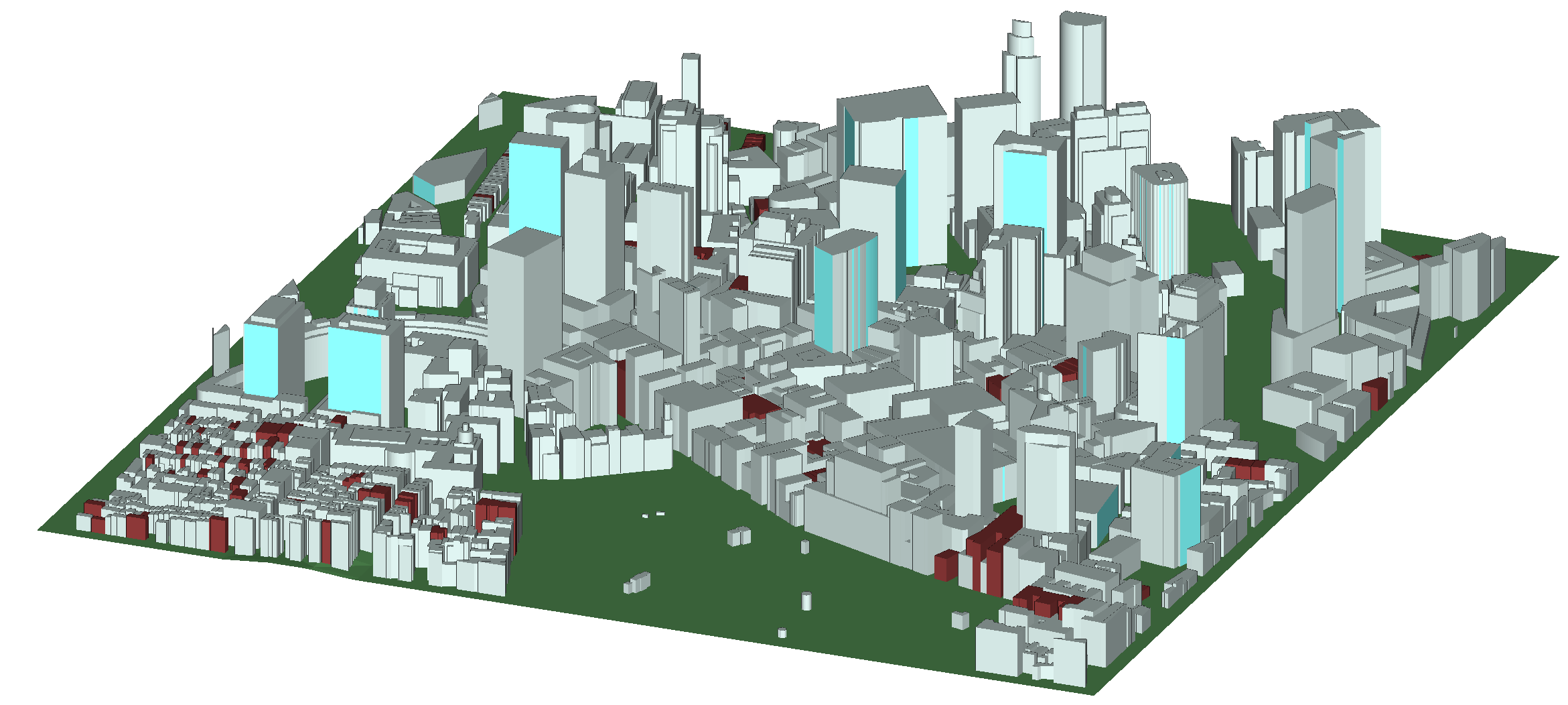}} 
\vspace{0.01\columnwidth}
\caption{3D representations of the four considered city sections: (a) Tokyo, (b) Beijing, (c) London, (d) Moscow, and (e) Boston.}
\label{fig:all_cities}
\end{figure*}

To generate data, we consider sections of five cities (Tokyo, Beijing, London, Moscow, and Boston) having varying sizes and distinct types of terrain, buildings, and foliage. Fig.~\ref{fig:all_cities} shows 3D representations of these city sections, whose blueprints are part of the Wireless Insite package. 
The number of deployed transmitters (\glspl{uav}) and receivers (gNBs) is detailed in Table~\ref{tab:city_param} for each of the cities. 
%All of these models include objects for the buildings, foliage, and terrain features of the respective city.

It is standard practice to differentiate channel models across environments, e.g., the 3GPP mmWave
%\footnote{\gio{[Minor]: I'd remove the word ``mmWave" because I think all mmWave calibrations are done for UMi. With UMa, you may not get sufficient coverage at 30 GHz.}} 
model provides separate parameter distributions for environments such as \emph{urban macro} and \emph{urban micro} \cite{3GPP38901}. In a data-driven approach,
environment-specific models can be created by partitioning the training data. % in our case the Tokyo-Beijing and the London-Moscow environments detailed in Section \ref{sec:remcom}.
In our case, we naturally define one distinct environment for each of the five represented city sections.

\begin{table*}[b!]
    \vspace{1mm}
    \caption{City sections and deployment parameters}
    \label{tab:city_param}
    \centering
    \begin{tabular}{l|c|c|c|c|c|}
    \cline{2-6}
    %\multirow{5}{*}{ } &  \cline{2-5}
    {}
    & \shortstack{Tokyo, Japan} 
    & \shortstack{Beijing, China} 
    & \shortstack{London, UK} 
    & \shortstack{Moscow, Russia}
    & \shortstack{Boston, USA} \\ \hline
    \multicolumn{1}{|l|}{Area ($m^2$)} &
    \shortstack{$1420 \times1440$} &
    \shortstack{$1650 \times1440$} &
    \shortstack{$1500 \times1480$} &
    \shortstack{$1440 \times1380$} &
    \shortstack{$1130 \times1220$} \\ \hline
    \multicolumn{1}{|l|}{Number of \glspl{uav}}
    & 140 & 120 & 120 & 160 & 138 \\ \hline
    \multicolumn{1}{|l|}{\shortstack{Number of standard gNBs}}
    & 220 & 180 & 122 & 200 & 95\\ \hline
    \multicolumn{1}{|l|}{\shortstack{Number of dedicated gNBs}}
    & 200 & 120 & 93 & 160 & 78\\ \hline
    \end{tabular}
\noindent
\end{table*}

%\begin{figure*}[ht!]
%\vspace{0.1\columnwidth}
%\centering
%\subfloat[Tokyo, Japan]{\includegraphics[width=0.45\columnwidth]{Figures/Tokyo.PNG}}\hspace{0.05\columnwidth}
%\subfloat[Beijing, China]{\includegraphics[width=0.45\columnwidth]{Figures/Beijing.PNG}}\hspace{0.05\columnwidth}
%\subfloat[London, UK]{\includegraphics[width=0.45\columnwidth]{Figures/London.PNG}}\hspace{0.05\columnwidth}
%\subfloat[Moscow, Russia]{\includegraphics[width=0.45\columnwidth]{Figures/Moscow.PNG}} 
%\caption{3D representations of the four city sections considered: (a) Tokyo, (b) Beijing, (c) London, and (d) Moscow. \\ \gio{Guys choose between the 2x2 and the 1x4 versions}}
%\label{fig:all_cities}
%\end{figure*}

%The unique features of each of these environments mean that a channel modeling structure such as our proposed two-stage generative model could capture subtleties that a standard channel model might not.
%This is more so the case when we begin to consider the deployment of theoretical aerial gNBs for the purposes of communicating and providing coverage to \glspl{uav}.

For our data production, as advanced in earlier sections,
%In our work, \glspl{uav} are distributed in open areas such as parks, and also positioned along street canyons as they might be during flight.
two distinct types of gNBs are manually placed:
\begin{itemize}[leftmargin=*]
    \item \emph{Standard gNBs}. These are
    placed on streets at a height of \SI{2}{m}, emulating typical locations
    for 5G microcells intended to serve ground users.  
    %We are interested in modeling the air-to-ground channels seen by gNBs at such locations, both to verify whether standard cells can serve UAVs and to understand the interference between UAV and ground-user transmissions therein.
    \item \emph{Dedicated gNBs}. These are located on rooftops, \SI{30}{m} above street level,
    meant to provide additional coverage to UAVs. % particularly at high altitudes.
\end{itemize}

Transmitting \glspl{uav}, for their part, are placed
at different horizontal locations in each environment at one of four possible
altitudes: 30, 60, 90 and \SI{120}{m}. 

In total, 58800 UAV-gNB links are created for the Tokyo environment, 36000 for Beijing, 25800 for London, 57600 for Moscow, and 23874 for Boston.
The Wireless InSite tool is then run to simulate the channel on every link, producing the path parameters $\nbx$ for each link.  
%(Although not used here, the ray tracing tool also returns the full route of each path including the scattering locations.)
All simulations are conducted at \SI{28}{GHz}.
% \red{William / Vasilii,  can you give the details on modeling of diffraction, number of reflections, permitivity assumptions of building surfaces. Maybe move to Fig. 2.  In full version, put back in the table with the parameters.}
The maximum number of reflections is set to six and the maximum number of diffractions is set to one, with
%Furthermore, the materials of different objects in the 3D model can be specified.
both ground and wall surfaces taken to be made of concrete with a permittivity of \SI{5.31}{F/m}.
The simulator provides the directions of arrival and departure, as well as the path losses and delays for each link.

The datasets thus gathered are utilized to train the model described in Sec.~\ref{sec:model}.

\section{Modeling Results}
\label{sec:results}

%\begin{figure*}

%\centering

%\begin{tabular}{cc}
%\textbf{Tokyo-Beijing} & 
%\textbf{London-Moscow} \\
%\includegraphics[width=0.9\columnwidth]{Figures/los_prob_tokyo_beijing.png}\hspace{0.1\columnwidth} & 
%\includegraphics[width=0.9\columnwidth]{Figures/los_prob_london_moscow.png} 
%\end{tabular}
%\caption{Conditional probability of a LOS link as
%a function of horizontal and vertical distance to the gNB for terrestrial and aerial types and
%for the two environments. 
%For each environment, the left-hand-side column is the 
%empirical distribution on the test data and the
%right-hand-side column is the probability forecast by the trained link-state predictor.}
%\label{fig:los_prob}
%\end{figure*}

%\subsection{Dataset partitions into Environments and Generalization Error}

%We evaluate the neural network models on the data in Section~\ref{sec:remcom}.
This section describes various features of the learned models,
and their ability to capture interesting wireless phenomena.
We also seek to evaluate the generalization ability of the models, meaning their ability to accurately describe the channel behavior in locations other than those in the training dataset. As mentioned in the introduction, this ability is a highly desirable attribute, and hence we test it extensively.

The links available for each environment are split, 75\% for training
and 25\% for testing. Models are then trained separately for each environment, which enables assessing the generalization ability in these two senses:
%We train models for the training data in each environment, BT and LM, separately.
%This division will enable us to evaluate the differences in these two types of cities.
%We will also be able to evaluate the models' generalization in two manners:
\begin{itemize}
\item \emph{Intra-environment}. The model trained on the 75\% training links of a specific dataset
%the Tokyo-Beijing (respectively London-Moscow) dataset
is evaluated on the 25\% test links of that same dataset.
%the Tokyo-Beijing (respectively London-Moscow) test data.
%Likewise, the model trained on the London-Moscow training dataset is evaluated on the London-Moscow test data.
This appraises the ability of the model to generalize to links in the same environment, but at new locations not seen during training.

\item \emph{Inter-environment}. The model trained on a specific dataset
%the Tokyo-Beijing dataset
is evaluated on another dataset. %the London-Moscow test data, and vice versa.
This serves to examine the model's ability to generalize to links in other environments.
\end{itemize}

All the implementations are based on Tensorflow 2.2; the code, datasets, and pre-trained models can be found in \cite{mmw-github}.
% \begin{figure}[h!]
%     \centering
%     \includegraphics[width=1.0\columnwidth]{Figures/los_prob_city.png}
%     \caption{LOS probability for ray-traced test data, averaged over the four possible UAV altitudes (30, 60, 90, and 120 m).}
%     \label{fig:los_prob_all}
% \end{figure}

% \begin{figure}[h!]
%     \centering
%     \includegraphics[width=1.0\columnwidth]{Figures/los_prob_city_model.png}
%     \caption{LOS probability generated by the model on the test locations, averaged over the four possible UAV altitudes (30, 60, 90, and 120~m).}
%     \label{fig:los_prob_all_model}
% \end{figure}

\subsection{LOS Probability}
\label{sec:link_state_eval}

\begin{figure*}
\centering
\subfloat[Ray-traced test data]{\includegraphics[width=0.7\columnwidth]{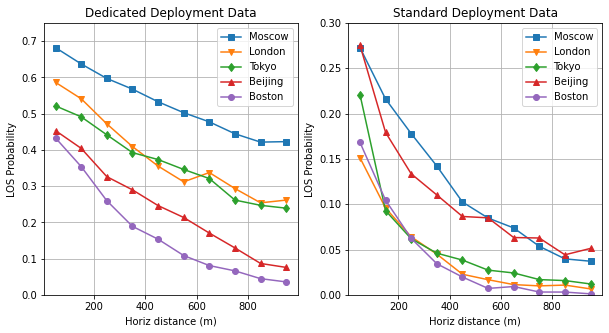}}
\hspace{0.005\columnwidth}
\subfloat[Proposed generative model]{\includegraphics[width=0.7\columnwidth]{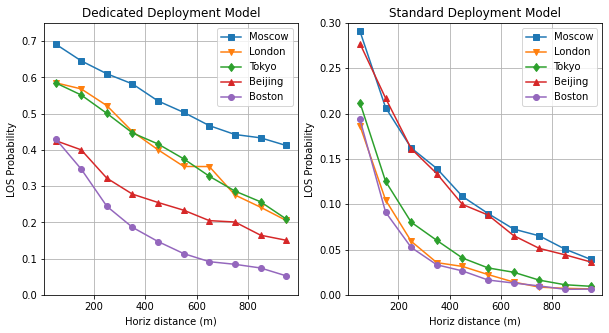}}\\
\caption{LOS probability computed for (a) the model and (b) from the model on the test locations, averaged over the four possible UAV altitudes (30, 60, 90, and 120~m).}
\label{fig:los_prob}
\end{figure*}
\begin{figure*}[h!]
\centering
\subfloat[Moscow]{\includegraphics[width=0.48\columnwidth]{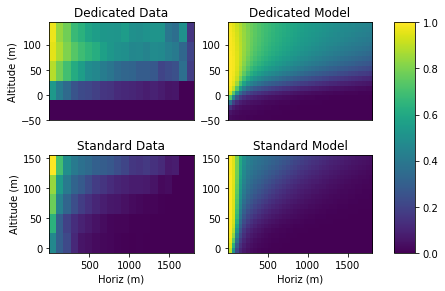}}
\hspace{0.005\columnwidth}
\subfloat[Tokyo]{\includegraphics[width=0.48\columnwidth]{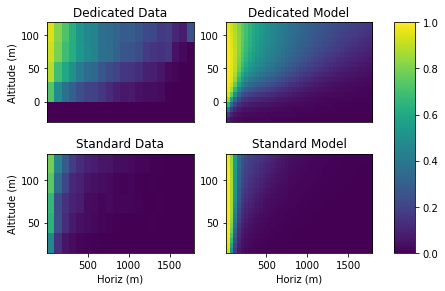}}\\
\caption{LOS probability for (a) Moscow, Russia and (b) Tokyo, Japan, parameterized by altitude and horizontal distance.}
\label{fig:los_prob_cities}
\end{figure*}
To illustrate the functioning of the link-state predictor,
Fig.~\ref{fig:los_prob} shows the probability of the
link being in the LOS state as a function of the horizontal
distance between UAV and gNB. Precisely, Fig.~\ref{fig:los_prob}a depicts the actual probabilities in the test data for each of the environments and Fig. \ref{fig:los_prob}b depicts the respective model predictions. In both cases, the results are averaged over the four possible UAV altitudes.

%For each environment, the bottom subfigures present the probability corresponding to terrestrial gNBs and the top subfigures present the probability corresponding to aerial gNBs, 
%while the left-hand-side columns show the actual values on the test data and the right-hand-side columns show the probability at the output of the trained link-state predictor.
% \begin{figure}
%     \centering
%     \includegraphics[width=1.0\columnwidth]{Figures/los_prob_moscow.png}
%     \caption{LOS probability for Moscow, parameterized by altitude and horizontal distance.}
%     \label{fig:los_prob_moscow}
% \end{figure}

% \begin{figure}
%     \centering
%     \includegraphics[width=1.0\columnwidth]{Figures/los_prob_tokyo.png}
%     \caption{LOS probability for Tokyo, parameterized by altitude and horizontal distance.}
%     \label{fig:los_prob_tokyo}
% \end{figure}

The link-state predictor is seen to accurately determine the trends in the test data for each of the environments and to reflect the very different behaviors of standard and dedicated gNBs.
We also observe interesting differences across environments. The \gls{los} probability is uniformly
higher in Moscow, both for standard and dedicated gNBs, consistent with the relatively shorter building therein.
Beijing, in turn, exhibits a relatively high LOS probability for standard gNBs yet a relatively low LOS probability for dedicated gNBs, a contrast that points to an abundance of both reflection opportunities and blockages.

Insights on the impact of the UAV altitude can be drawn from Figs. \ref{fig:los_prob_cities}a and \ref{fig:los_prob_cities}b, where again we see the excellent match between the test data and the model predictions thereon. Dedicated gNBs can provide substantially higher probabilities of LOS coverage at long horizontal distances provided the UAV is high enough. 

%\angel{How can we say that if the results are not parameterized by altitude? William: I think I may include a single plot of the LOS probability as parameterized by horizontal distance and altitude (as we had previously), but for only one city. What do you think?}\angel{Yes, if we have one plot that clearly supports this observation, we should include it. Otherwise we'll have to tone down the comment.}
In contrast, standard gNBs tend to be far more limited in terms of horizontal coverage.
%These observations are supported by the data in Figs. \ref{fig:los_prob_moscow} and \ref{fig:los_prob_tokyo}, and the behavior is predicted by the proposed generative model.

%at greater distances, both for terrestrial and aerial gNBs. This is consistent with the relatively smaller building heights therein.
We will see next how all of the above has a significant impact in other features such as the path loss and path angular distributions.

%\begin{figure}
%\centering
%\includegraphics[width=0.999\columnwidth]{
%Figures/path_loss_cdf_tb.png}
%\caption{Intra-environment prediction of omnidirectional
%path losses: \emph{``TokBei data''} is 
%the CDF for links in the test dataset of the
%Tokyo-Beijing environment;
%\emph{``TokBei model''}
%is the CDF of values generated by the model trained on the same Tokyo-Beijing dataset with the same link conditions as the test data. In both %cases, the path loss is only 
%plotted for the links in {\tt LOS} and {\tt NLOS} states, with {\tt NoLink} cases excluded.}
%\label{fig:omni_path_loss_tb}
%\end{figure}

\subsection{Path Loss: Intra-Environment Evaluation}
%\subsection{Intra-Environment Evaluation of Predicting the Path Loss with Omnidirectional Antennas}
\begin{figure*}[b]
\centering
\subfloat[Boston, USA]{\includegraphics[width=0.49\columnwidth]{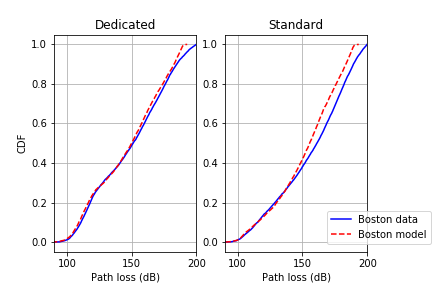}}
\subfloat[Moscow, Russia]{\includegraphics[width=0.49\columnwidth]{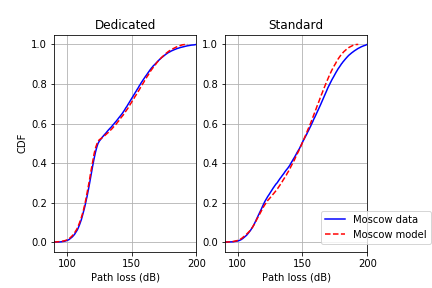}}
\caption{CDF of the path loss for (a) Boston and (b) Moscow.}
\label{fig:cities_pl}
\end{figure*}
We now turn to evaluating the accuracy of the rest of the
parameters. Fundamentally, we want to measure
how close the distribution of the trained generative
model in \eqref{eq:xgenuz} is to the observed conditional
distribution of the test data itself.
To this end, let $(\nbu_i,\nbx_i)$, $i=1,\ldots,N_{\rm ts}$ be
the test samples, each containing a link
condition, $\nbu_i$, and its corresponding path parameters, $\nbx_i$.
To evaluate how closely the learned model fits this test data,
for each sample we can compute some 
statistic $\phi(\nbu_i,\nbx_i)$ that is of relevance.
As an example of statistic, we compute the path loss experienced by UAVs and gNBs equipped with omnidirectional antennas, deferring to later in the paper the consideration of directivity.
%directional antennas are entertained later in the paper.

Using the same conditions $\nbu_i$ in the test data,
we generate a sample $\nbx^{\rm rnd}_i=g(\nbu_i,\nbz_i)$
from the trained generative model and some random $\nbz_i$.
We can then compute 
$\phi(\nbu_i, \nbx^{\rm rnd}_i)$
and compare its CDF with that of the actual $\phi(\nbu_i, \nbx_i)$.

We first evaluate the intra-environment accuracy
of the omnidirectional path loss predictions.
Fig.~\ref{fig:cities_pl} shows the CDF of path losses for the test data of a couple of environments %(curve labeled \emph{``TokBei data''}).
alongside the
CDF of path losses generated by the trained model using the same condition values as the test data.  
An excellent match is observed for both standard and dedicated gNBs.
In particular, the trained generative model is able to capture the multi-slope behavior that arises in some environments due to the mixture of LOS/NLOS links. 

%\footnote{\gio{Can we say anything about the shape of the CDFs in Fig.~\ref{fig:omni_path_loss}? Both seem to have a dual slope, which could correspond to LoS/NLoS links. E.g., for Terrestrial 30\% of all links are in LoS and have small path loss ($<110$dB), whereas for Aerial 65\% of all links are in LoS with small path loss ($<110$dB). This is just a conjecture.}}

%\footnote{\gio{Fig.~\ref{fig:omni_path_loss}: better not relying on colors only, but also on different line style or markers. Also the resolution of the figure is low, do we have .eps?}}
\begin{figure*}[b]
%\vspace{0.1\columnwidth}
\centering
\subfloat[Beijing model on Moscow data.]{\includegraphics[width=0.49\columnwidth]{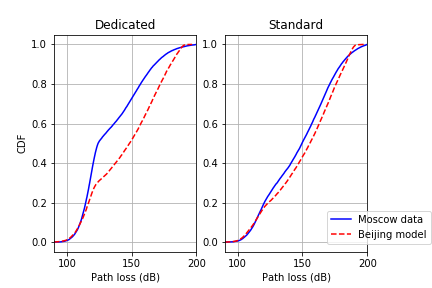}}
\subfloat[London model on Tokyo data.]{\includegraphics[width=0.49\columnwidth]{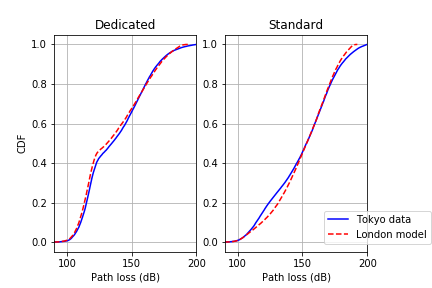}}
\caption{Inter-environment comparisons for (a) a model that fails to accurately predict the path loss on an environment different than the training one, and (b) a model that does accurately make that prediction.} 
\label{fig:inter-city_pl}
\end{figure*}

%\begin{figure}
%\centering
%\includegraphics[width=0.999\columnwidth]{
%Figures/path_loss_cdf_tb_lm.png}
%\caption{Inter-environment prediction of omnidirectional 
%path losses: \emph{``TokBei data''} and
%\emph{``TokBei model''} are the CDFs
%from Fig.~\ref{fig:omni_path_loss_tb};
%\emph{``LonMos model''} is the 
%CDF of values
%generated by the model trained on the London-Moscow
%dataset;
%\emph{``LonMos path mod only''} is the CDF when using
%the path generator trained from the London-Moscow data,
%but with the actual link state (LOS, NLOS and NoLink) 
%as opposed to the model-generated one.
%}
%\label{fig:omni_path_loss_tb_lm}
%\end{figure}

\subsection{Path Loss: Inter-Environment Evaluation}
%\subsection{Inter-Environment Evaluation of the Omni-Directional Path Loss}

%The previous sub-section assessed the ability of the model to predict the omni-directional path loss on links in the \emph{same} environment, specifically links in Tokyo and Beijing.  As discussed earlier, an important question is how well do models trained in one environment describe links in a \emph{different} environment.

%To test the inter-environment generalization ability

%Fig.~\ref{fig:omni_path_loss_tb_lm} reproduces again, as baselines, the CDF of omnidirectional path losses for the test data in the Tokyo-Beijing dataset, and from the model trained on that same environment. As observed in Fig.~\ref{fig:omni_path_loss_tb}, these CDFs match well, indicating that the modeled trained in a given environment can predict the behavior of new links in that same environment.

Next, we gauge the model's ability to make predictions on an environment after having been trained on a different one. Presented in Fig. \ref{fig:inter-city_pl} are two sets of contrasting such results.
On the left-hand side we have the CDF of the path loss on the Moscow test data as predicted by a model trained with the Beijing dataset. For standard gNBs the match is satisfactory, indicating similarity in the respective propagation mechanisms for those gNBs, chiefly reflections. For dedicated gNBs, conversely, the Beijing model largely overstates the Moscow path loss, pointing to important discrepancies in the degree of blockage between the two environments. These observations are fully consistent with those made in Section \ref{sec:link_state_eval} for the LOS probabilities in Moscow and Beijing.
On the right-hand side of the figure, the same exercise is repeated for a model trained with London data and tested in the Tokyo environment, and in this case the agreement is excellent for both standard and dedicated gNBs.
We thus see how the proposed methodology enables 
assessing the inter-environmental generalizability
of models, which turns out to depend not only on the environments but further on the type of gNB.
The similarities and discrepancies thereby revealed are valuable and highly non-obvious from a visual inspection of Fig. \ref{fig:all_cities}.

\begin{figure*}[th!]
%\vspace{0.1\columnwidth}
\centering
\subfloat[Tokyo, Japan]{\includegraphics[width=0.25\columnwidth]{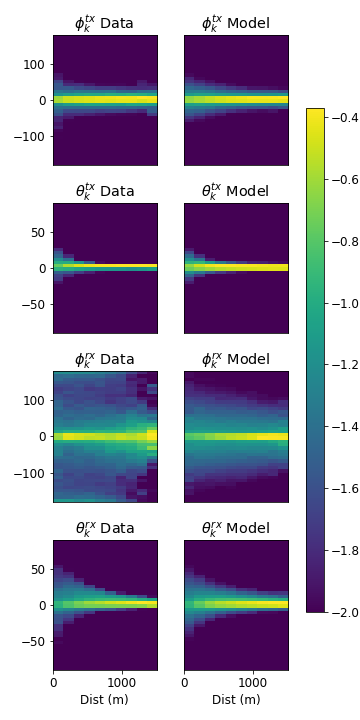}}\hspace{0.05\columnwidth}
\subfloat[Beijing, China]{\includegraphics[width=0.25\columnwidth]{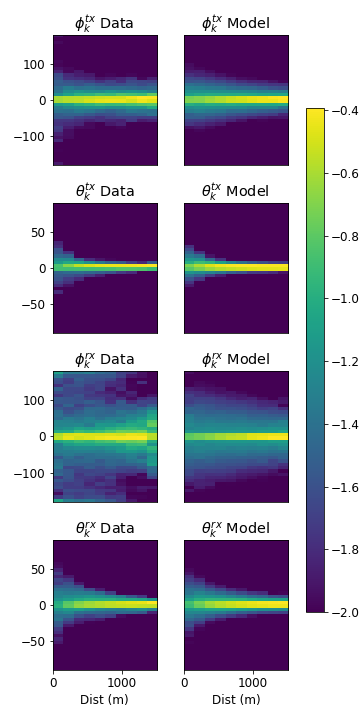}}\hspace{0.05\columnwidth}
\subfloat[London, UK]{\includegraphics[width=0.25\columnwidth]{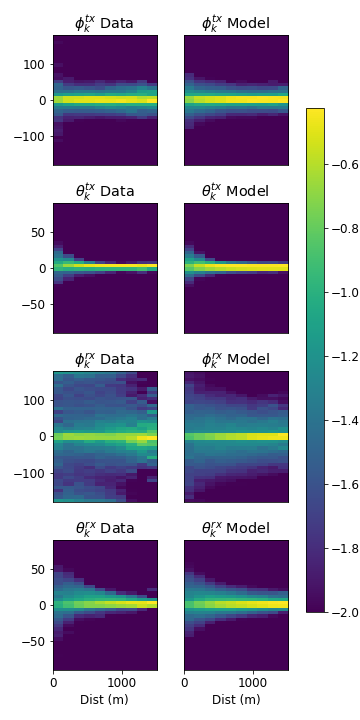}}\\\vspace{0.05\columnwidth}
\subfloat[Moscow, Russia]{\includegraphics[width=0.25\columnwidth]{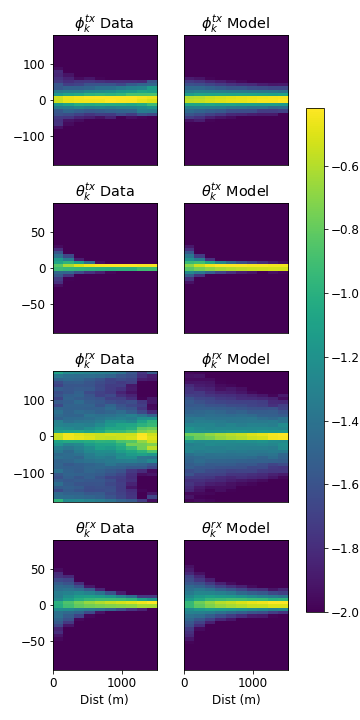}} \hspace{0.05\columnwidth}
\subfloat[Boston, USA]{\includegraphics[width=0.25\columnwidth]{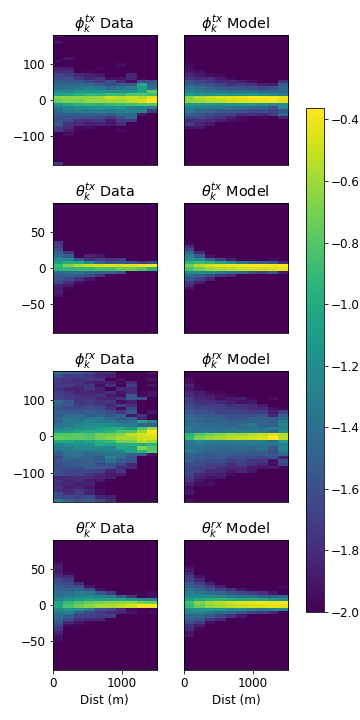}} 
\caption{Distribution of angles, averaged over the four UAV altitudes, for (a) Tokyo, (b) Beijing, (c) London, (d) Moscow, and (e) Boston.}
\label{fig:angle_dist}
\end{figure*}

\subsection{Angular Distribution}
Let us now turn to the path angles.
Fig.~\ref{fig:angle_dist} plots the distribution of those angles as a function of the 3D
distance between the UAV and gNB.
The distribution is computed over all the paths within \SI{30}{dB} of the strongest
path within each link for all the links in the test dataset, and it is averaged over the four possible UAV altitudes.
(For the sake of readability, the links to standard and dedicated gNBs are combined, but respective plots for the standard and dedicated gNBs, or plots to separately observe the effects of elevation and horizontal distance, could just as well be produced from the model.)

Each row in Fig.~\ref{fig:angle_dist} shows
the distribution of one of the four angles,
$\phi^{\rm rx}_k,\theta^{\rm rx}_k,\phi^{\rm tx}_k,\theta^{\rm tx}_k$, relative to the LOS direction
(even when the LOS path is blocked).
For each environment,
the left-hand-side column is the distribution for the test data whereas the right-hand-side column
is the its counterpart generated by the learned model.

The model matches very well the actual angular distribution in the test data.
In particular, it captures an important phenomenon: for all distances
and angles, the NLOS paths tend to be angularly
close to the LOS direction.
Moreover, the angular spread decreases %---particularly at the gNB side---
as the UAV and gNB are further apart.
This behavior makes intuitive sense in that, as the UAV
pulls away from the gNB, there is less local scattering to create angular dispersion.
%\footnote{\gio{A limitation of this is that we cannot see the effect of the UAV height.}}
Consistent with this, the scattering is much wider at the gNB end of the links.

%It can also be observed that there is much greater scattering at the gNB side (the angles of arrival). This is due to the fact that, in many test locations, the UAV is elevated with minimal local scattering.

\subsection{SNR Predictions}
\begin{figure*}[th]
%\vspace{0.1\columnwidth}
\centering
\subfloat[Tokyo, Japan]{\includegraphics[width=0.45\columnwidth]{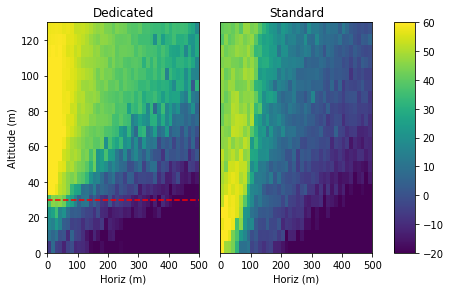}}\hspace{0.05\columnwidth}
\subfloat[Beijing, China]{\includegraphics[width=0.45\columnwidth]{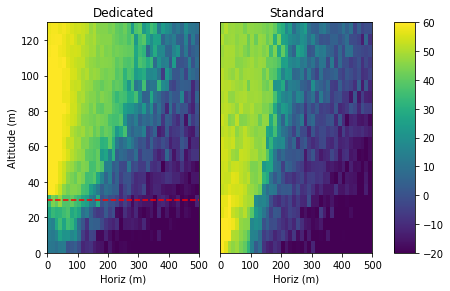}}\\
\subfloat[London, UK]{\includegraphics[width=0.45\columnwidth]{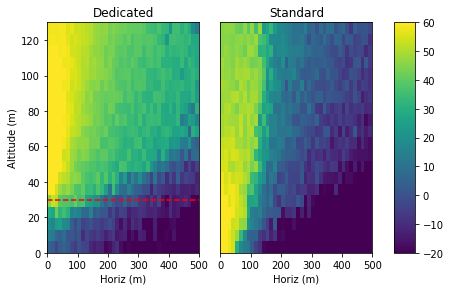}}\hspace{0.05\columnwidth}
\subfloat[Moscow, Russia]{\includegraphics[width=0.45\columnwidth]{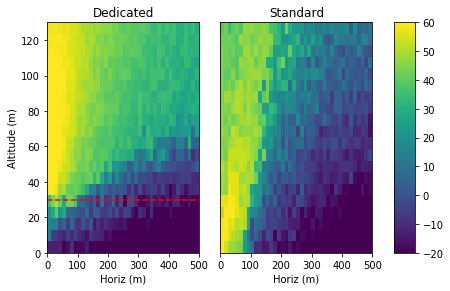}} \hspace{0.05\columnwidth}
\subfloat[Boston, USA]{\includegraphics[width=0.45\columnwidth]{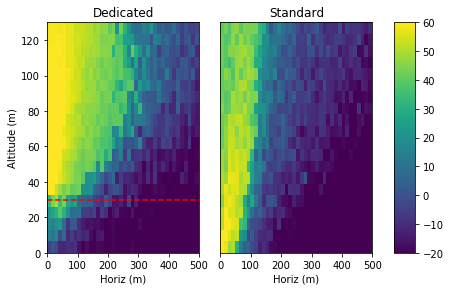}} 
\caption{Median predicted local-average SNR as a function of the UAV position for (a) Tokyo, (b) Beijing, (c) London, (d) Moscow, and (e) Boston. The horizontal lines indicate the altitude of the dedicated gNBs. }
\label{fig:cities_snr}
\end{figure*}

We finalize by demonstrating a specific application enabled by the generative model.
Specifically, we compute the 
predicted uplink (UAV to gNB) local-average SNR as a function of the UAV position
in the single-cell scenario described in Table~\ref{tab:link_budget_param},
which is consistent with current \SI{28}{GHz} 5G deployments \cite{xia2019uav}.
Such uplink SNR is of particular interest since this is usually the power-limited link direction, and the one envisioned to support high-bit-rate applications \cite{3GPP22125,3GPP22829}.
A gNB is located at $(0,0,h)$ with $h=2$~m and $h=30$~m in the standard and dedicates cases, respectively. In the standard case, the gNB features three sectors with a half-power beamwidth of $90^\circ$ per sector and a $100^\circ$ downtilt (relative to vertical), as customary to serve ground users. Hence, the connections from UAVs to standard gNBs must necessarily be through sidelobes or reflected paths \cite{3GPP36777,GerGarGal2018}.
In the dedicated case, the gNB is single-sectored with an upward-facing array intended for aerial coverage.
%\begin{figure}[t]
%\centering
%\includegraphics[width=\columnwidth]{Figures/snr_dist.png}
%\caption{Median SNR predicted
%by the model as a function of the 
%horizontal and elevation position of the UAV
%for the Tokyo-Beijing environment.
%Details are provided in Table~\ref{tab:link_budget_param}.
%The red dotted line indicates the height of the aerial gNB.} % \gio{typo ``Terestrial''}
%\label{fig:snr_dist}
%\end{figure}
The UAV, equipped with a single array at its bottom, designed for lower-hemisphere coverage \cite{xia2019uav}, is at
$(x,0,z)$ with $x \in [0,500]$~m
and $z \in [0,130]$~m.  
For each UAV position and gNB type
(standard or dedicated), 
100 channels realizations are generated by the 
model and used to compute the local-average SNR \cite{lozano2012yesterday}.
%We apply a model learned from the Tokyo-Beijing data, yet  a similar plot could be generated from the London-Moscow dataset as well.
%
%From the channel paths and the link budget values in Table~\ref{tab:link_budget_param}, which are consistent with current \SI{28}{GHz} 5G deployments \cite{xia2019uav}, the local-average wideband SNR is computed.
Fig.~\ref{fig:cities_snr} plots the median such SNR.

\begin{table}[bh!]
    \caption{Uplink single-cell simulation parameters.}
    \label{tab:link_budget_param}
    \centering
    % 1.5cm, 6.2cm spacing for double column
    \begin{tabulary}{\columnwidth}{ |p{3.4cm} | p{7cm} | }
    \hline
    \text{Item} & \text{Value} \\ \hline\hline
    \multirow{2}{*}{Spectrum}		& Carrier frequency: 28~GHz \\ \cline{2-2}
		& Bandwidth: 400 MHz ($4 \times 100$ MHz aggregation) \\ \hline
    gNB height		& Standard: 2~m; Dedicated: 30~m \\ \hline
    \multirow{2}{*}{Array size}		& UAV: $N_{\rm UAV}$ = 16 
    ($4 \times 4$ UPA) \\ \cline{2-2}	& gNB: $N_{\rm gNB}$ = 64 
    ($8 \times 8$ UPA) \\ \hline
    Antenna spacing & Half-wavelength \\ \hline
    %PA efficiency (\%) & 20 &  
    %Values from \cite{sadhu201728}
    %with backoff
    %\tabularnewline \hline
    %UAV PA power (W)   & 1.0 &
    %\tabularnewline \hline
    \multirow{3}{*}{\shortstack{Array vertical orientation}}	& UAV: $180^{\circ}$ $\downarrow$ lower hemisphere coverage \cite{xia2019uav} \\ \cline{2-2}	& Standard gNB: $100^{\circ}$ $\searrow$ ground coverage, 3 sectors \\ \cline{2-2} & Dedicated gNB: $0^{\circ}$ $\uparrow$ upper hemisphere coverage  \\ \hline
    Transmit power	& UAV: 23 dBm \\ \hline 
    Losses		& 6 dB including noise figure \cite{ChenRfic2018,Garg17RTPS} \\ \hline 
	\end{tabulary}
\end{table}

The experiment shows how the SNR at any location within the environment can be predicted from
the model and the specifics of the setting (arrays, powers, and the other details in Table \ref{tab:link_budget_param}).  
The dedicated gNBs provide much better
coverage at large horizontal distances, yet standard gNBs can provide solid coverage
when the horizontal distance is small (below roughly $100$~m).  This coverage from standard gNBs
is rather surprising: complying with 3GPP specifications \cite{3GPP38901}, standard gNBs have downtilted antennas with a \SI{30}{dB} front-to-back gain ratio, which 
hinder the connectivity from direct vertical paths.
However, the learned model captures reflections
from neighboring buildings within the antenna
beamwidth, and the simulations show that these reflected paths 
do enable coverage.

\section{Bechmarking Against 3GPP Models}
\label{subsec:compare_3gpp}

%We now turn to a more quantitative approach to understanding the performance of our proposed two-stage architecture.
%This section will also detail an effort to compare our generative scheme to that of both a default and re-parameterized standard model, namely the \gls{3gpp} model.
%Note that we will use specifically the UMi-AV scenario, since it is the channel model that \gls{3gpp} developed for aerial vehicles.

To complete the test-drive of the proposed generative model, it is fitting to benchmark it against the alternatives offered by existing standards.  We focus here
on the 3GPP UMi-AV (urban micro with aerial vehicles) scenario \cite{3GPP38901,3GPP36777},
which is closest to our work.  
This model is suitable for mmWave frequencies up to \SI{100}{GHz}, but its parameters are calibrated for UAVs---meaning users at altitudes above 22.5 m---only
below \SI{6}{GHz}.
In order to provide a fair benchmark for our proposed architecture, we refit those parameters with the data for each of our environments,
and restrict the comparisons to standard gNBs.

%and compare the accuracy of the fitted 3GPP model against the proposed neural network based method.

%This process consists of first determining what constant parameters characterize the UMi-AV channel model, then finding new values for those constants that allow the model to best describe the multi-path components of each city. 
%Critically, a new set of parameters will be determined for each city, allowing for maximum flexibility of the formerly generalized standard model.

%\angel{Let's indicate that, for the path loss at least, we're extrapolating a sub-6 GHz model.}
%\angel{William, let's explain here what we're going to do in this section, so the term "re-parameterization" doesn't hit the reader as a surprise when it appears.}

\subsection{LOS Probability}
\label{subsubsec:3gpp_los}
We first examine the probability of \gls{los}, $P_{\rm LOS}$.
The 3GPP model takes in several parameters such as the heights of transmitter and receiver and, as well as their horizontal distance, $d_{\rm 2D} = \sqrt{d^2_x + d^2_y}$ \cite{3GPP38901}. 
If the \gls{uav} height $h$ is between \SI{1.5}{\meter} and \SI{22.5}{\meter}, then 
\begin{equation}
    P_{\rm LOS} = 
    \begin{cases}
        1 & d_{\rm 2D} \leq \alpha_1 \\
        \frac{\alpha_1}{d_{\rm 2D}}+e^{-\frac{d_{\rm 2D}}{\alpha_2}}\left( 1-\frac{\alpha_1}{d_{\rm 2D}}\right) & \alpha_1 \leq d_{\rm 2D}
    \end{cases} 
    \label{eq:plos_low}
\end{equation}
whereas, if $h>$ \SI{22.5}{\meter},
\begin{equation}
    P_{\rm LOS} = 
    \begin{cases}
        1 & d_{\rm 2D} \leq d_1 \\
        \frac{d_1}{d_{\rm 2D}} + e^{-\frac{d_{\rm 2D}}{p_1}}\left( 1 - \frac{d_1}{d_{\rm 2D}}\right) & d_{\rm 2D} > d_1
    \end{cases}
    \label{eq:plos_high}
\end{equation}
with
\begin{equation}
\begin{split}
    & p_1 = \alpha_3\log_{10}(h)+\alpha_4 \\
    & d_1 = \max(\alpha_5\log_{10}(h)+\alpha_6,\alpha_1) .
\end{split}
\label{eq:plos_vars}
\end{equation}
The values for the parameters, which in the 3GPP model \cite{3GPP38901} are
\begin{align}
\bm{\alpha}_{\rm LOS} & = [\alpha_1, \alpha_2, \alpha_3, \alpha_4, \alpha_5, \alpha_6] \nonumber \\ \label{eq:alpha_nom_plos}
  & = [18, 36, 294.05, -432.94, 233.98, -0.95],
\end{align}
are herein refitted for each environment. % via maximum likelihood.
Specifically, for each link we specify the set of condition variables
\begin{equation} \label{eq:input_3gpp_plos}
    \mathbf{u} = \left[\log_{10}(h),d_{\rm 2D},h, h_{\rm gNB}\right],
\end{equation}
and a binary label $y = 1$ if the link is LOS
and $y=0$ otherwise.
The 3GPP model
can be viewed as a function
\begin{equation}
    P_{\rm LOS} = P(y=1|\mathbf{u}) = 
    g_{\rm LOS}(\mathbf{u}, \bm{\alpha}_{\rm LOS}),
\end{equation}
mapping the condition vector
$\mathbf{u}$ to the LOS probability.
From the links $(\mathbf{u}_i,y_i)$ on a given environment,
the parameters $\bm{\alpha}_{\rm LOS}$ can be found
by minimizing the binary cross entropy (BCE),
\begin{align} \label{eq:plos_bce}
   J(\bm{\alpha}_{\rm LOS}) & = -\sum_i \Big[ y_i\log( g_{\rm LOS}(\bm{u}_i,\bm{\alpha}_{\rm LOS} ))  \\
    & \quad + (1-y_i)\log(1 - g_{\rm LOS}(\bm{u}_i, 
    \bm{\alpha}_{\rm LOS})) \Big].
\end{align}
This minimization, which is tantamount to a maximum likelihood
estimation of $\bm{\alpha}_{\rm LOS}$, is performed via stochastic gradient descent
(see Table~\ref{table:3gpp_nn_param}).
\begin{table}[h]
    %\vspace{1mm}
    \caption{3GPP Refitting Optimization}
    \label{table:3gpp_nn_param}
    \centering
    \begin{tabular}{l|c|c|}
    \cline{2-3}
    {} &
    \shortstack{$P_{\rm LOS}$} & \shortstack{Path Loss} \\ \hline
    \multicolumn{1}{|l|}{Number of inputs} &
    5 & 6 \\ \hline
    \multicolumn{1}{|l|}{Number of parameters} &
    6 & 19 \\ \hline
    \multicolumn{1}{|l|}{Optimizer} &
    \multicolumn{2}{c|}{Adam} \\ \hline
    \multicolumn{1}{|l|}{Loss function} &
    Binary cross entropy & Mean-squared error \\ \hline
    \multicolumn{1}{|l|}{Learning rate} &
    \multicolumn{2}{c|}{$10^{-3}$} \\ \hline
    \multicolumn{1}{|l|}{Epochs} &
    \multicolumn{2}{c|}{50} \\ \hline
    \multicolumn{1}{|l|}{Batch size} &
    \multicolumn{2}{c|}{128} \\ \hline
    \end{tabular}
\noindent
\end{table}
A distinct set of refitted parameters is obtained for each of the environments in
Fig.~\ref{fig:all_cities}, with the imposition that those parameters are within a multiplicative interval $[0.01,10]$ of the nomimal 3GPP values in \eqref{eq:alpha_nom_plos}
to prevent overfitting.

Our proposed generative approach can now be validated against 
the default \gls{3gpp} model %(with its nominal parameters)
and its refitted version.
%First, we take the test data points (links not used in training) in each environment.  
The horizontal distance, $d_{\text{2D}}$, and the vertical distance, $d_z$, are binned into sections of \SI{20}{m} and \SI{5}{m}, respectively. From a
histogram of the test links' LOS condition over the bins, the empirical $P_{\rm LOS}$ of the test data is obtained and contrasted with the prediction from the three models.
%The horizontal and vertical axes of the test data points are divided into a grid that is uniformly spaced in \SI{20}{m} increments in the horizontal plane, and in \SI{5}{m} increments in elevation. The spaces of the grid serve as bins where $P_{\rm LOS}$ of the test links are contrasted with the predicted probability from the three models.

Table~\ref{tab:plos_error} shows the mean absolute
error of the LOS probability over the grid.
We observe that the refitted 3GPP model is significantly better than its default form and that our proposed approach,
with minimal prior structure, performs similarly or better---sometimes markedly---than even the refitted
3GPP model on every environment.
%For Beijing and London specifically, the trained \gls{3gpp} model matches or just narrowly outperforms our proposed model. 
%Our proposed link-state predictor, however, has a lower error than the trained \gls{3gpp} model for the other cities, sometimes significantly.  

\begin{table*}[h!]
    \vspace{1mm}
    \caption{$P_{\rm LOS}$: Mean Absolute Error}
    \label{tab:plos_error}
    \centering
    \begin{tabular}{l|c|c|c|c|c|}
    \cline{2-6}
    {}
    & \shortstack{Tokyo, Japan} 
    & \shortstack{Beijing, China} 
    & \shortstack{London, UK} 
    & \shortstack{Moscow, Russia}
    & \shortstack{Boston, USA} \\ \hline
    \multicolumn{1}{|l|}{Default 3GPP Model}
    & \cellcolor{red!25}0.272 & \cellcolor{red!25}0.247 & \cellcolor{red!25}0.303 & \cellcolor{red!25}0.180 & \cellcolor{red!25}0.336 \\ \hline
    \multicolumn{1}{|l|}{Refitted 3GPP Model}
    & 0.040 & \cellcolor{green!25}0.056 & \cellcolor{green!25}0.057 & 0.058 & 0.047\\ \hline
    \multicolumn{1}{|l|}{Proposed Generative Model}
    & \cellcolor{green!25}0.036 & 0.058 & \cellcolor{green!25}0.057 & \cellcolor{green!25}0.034 & \cellcolor{green!25}0.041\\ \hline
    \end{tabular}
\noindent
\end{table*}

\subsection{Path Loss}
\label{subsubsec:3gpp_pl}
\begin{figure}[b]
    \centering
    \includegraphics[width=0.49\columnwidth]{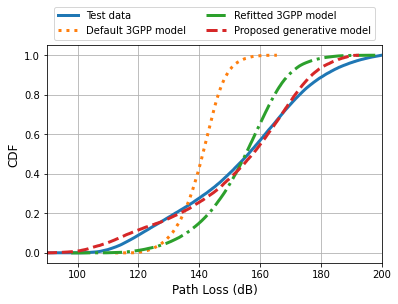}
    \caption{Path loss fitting for Tokyo, Japan.} 
    \label{fig:model_compare_pl_tokyo}
\end{figure}
We employ a similar strategy to refit the \gls{3gpp} path loss model. % in \cite[Table B-1]{3GPP36777}.
It is important to note that, unlike with $P_{\rm LOS}$, which depends exclusively on the geometry, the path loss is frequency-dependent. 
%We proceed with the acknowledgement that the 3GPP model structure was devised from measurements at sub-\SI{6}{GHz} frequencies.
% Thus, we have path loss nominal parameters 
% \begin{equation}
% \begin{split}
%     & \bm{\alpha}_{nom,los} = [32.4, 21, 20, 32.4, 40, 20, -9.5, 30.9, 22.5, -0.5, 20] \\
%     & \bm{\alpha}_{nom,nlos} = [35.3, 22.4, 21.3, -0.3, 32.4, 43.2, -7.6, 20]
% \end{split}
% \end{equation}

Separately for the LOS and NLOS cases, the 3GPP
model accepts as input the condition vector in (\ref{eq:input_3gpp_plos})
%\begin{equation}
%    \mathbf{x} = \left[\log_{10}(h_{UT}),d_{2D},h_{UT}, h_{gNB} \right],
%\end{equation}
and outputs a predicted path loss as some function,
\begin{equation}
    \mathrm{PL} = g_{\rm PL}(\bm{x}, \bm{\alpha}_{\rm PL} ),
\end{equation}
for specific parameters $\bm{\alpha}_{\rm PL}$
whose nominal details are given in \cite[Table B-1]{3GPP36777}.
We refit the model, this time with a mean-squared error
criterion (see Table~\ref{table:3gpp_nn_param} for details).

An example is presented in Fig.~\ref{fig:model_compare_pl_tokyo}, which depicts
the CDFs of path losses for the test data, the default \gls{3gpp} model, the refitted \gls{3gpp} model, and our proposed approach for Tokyo specifically.
While the refitted 3GPP model performs decidedly better than the default one, our proposed approach best approximates the distribution of the actual data.
%A similar behavior is observed for the other environments.
To quantify the differences among the distributions in this and the other environments, we invoke the 
Wasserstein-1 distance \cite{ruschendorf1985wasserstein}.
%between the empirical path loss values in the test data and the samples of the path losses generated from default 3GPP model, re-parametrized 3GPP model and the proposed neural network.
For two distributions $P$ and $Q$, the Wasserstein distance equals
\begin{equation}
    W(P,Q) = \max_{f} \Big[ \mathbb{E}(f(X)|X \sim P) - 
            \mathbb{E}(f(X)|X \sim Q)\Big],
\end{equation}
where the maximization is over all Lipschitz functions
satisfying $\|\nabla f(x)\| \leq 1$ $\forall x$. 
This metric is commonly used to train GANs
\cite{gulrajani2017improved} and, for scalar random variables,
it can be computed efficiently as
the integrated difference in CDFs
\cite{vallender1974calculation}.  Table~\ref{tab:pl_error}
shows the Wasserstein distance between the test data and the various models, with the propounded approach outperforming the rest in every environment.

\begin{table*}[h]
    \vspace{1mm}
    \caption{Path loss: Wasserstein-1 distance to test data distribution (dB)}
    \label{tab:pl_error}
    \centering
    \begin{tabular}{l|c|c|c|c|c|}
    \cline{2-6}
    {}
    & \shortstack{Tokyo, Japan} 
    & \shortstack{Beijing, China} 
    & \shortstack{London, UK} 
    & \shortstack{Moscow, Russia}
    & \shortstack{Boston, USA} \\ \hline
    \multicolumn{1}{|l|}{Default 3GPP Model}
    & \cellcolor{red!25}15.8 & \cellcolor{red!25}18.8 & \cellcolor{red!25}17.8 & \cellcolor{red!25}15.5 & \cellcolor{red!25}21.4 \\ \hline
    \multicolumn{1}{|l|}{Refitted 3GPP Model}
    & 10.7 & 14.8 & 12.3 & 14.3 & 14.9\\ \hline
    \multicolumn{1}{|l|}{Proposed Generative Model}
    & \cellcolor{green!25}6.70 & \cellcolor{green!25}2.22 & \cellcolor{green!25}2.95 & \cellcolor{green!25}2.49 & \cellcolor{green!25}3.42\\ \hline
    \end{tabular}
\noindent
\end{table*}

\section{Conclusion}
\label{sec:conclusion}

Generative NNs are a fitting engine for statistical channel modeling in complex settings such as those encountered in mmWave UAV communication.
Provided that abundant data is available, generative NNs are perfectly equipped to learn intricate probabilistic relationships and then produce parameters distributed accordingly. The only assumption is the choice of the parameters themselves, which can rest on basic principles of radio propagation.

%This paper has validated the methodology for an air-to-ground channel, in itself a prime example of complex setting, and specifically for an urban environment at mmWave frequencies. 
The proposed generative model, publicly available \cite{mmw-github}, has been shown to learn effectively and it can therefore be calibrated for any desired operating frequency, type of deployment, and environment, for which representative data is available.
The model can then capture any dependencies present in the data. In current standard-defined aerial channels, for instance, the distributions from which the angles of the multipath components are drawn do not depend on the distance; in contrast, and as intuition would have it, our model indicates a progressive narrowing of these distributions over distance.

In closing, we recall that, while the model has proved its ability to learn and to made interesting predictions driven by ray-tracing data, the ultimate objective is to drive it with empirical data. For this purpose, a measurement collection campaign is underway. 
%Likewise, benchmarking against standard-defined models should take place whenever such models become available for mmWave aerial channels.

%The generality of any data-based model hinges on the representativity of the data itself, and learning-based approaches are no exception. This aspect is inherently two-sided. On the one hand, generality is a desirable attribute for the purposes of research, development, algorithm comparison, and even standardization. On the other hand, site-specificity, i.e., lack of generality, is decidedly a virtue when it comes to actual deployments.

%Ideally then, the model should be able to train to varying degrees of generality.

%\footnote{\gio{Shall we summarize takeaways? Shall we mention future work, e.g., SINR accounting for interference, aerial cell deployment study to meet coverage requirements, etc.}}

%\section*{Acknowledgements}

%W.~Xia, S.~Rangan, and M.~Mezzavilla were supported by NSF grants  1302336,  1564142,  1547332, and 1824434,  NIST, SRC, and the industrial affiliates of NYU WIRELESS. A.~Lozano and G.~Geraci were supported by ERC grant 694974, by MINECO's Project RTI2018-101040, by ICREA, and by the Junior Leader Fellowship Program from ``la Caixa" Banking Foundation. V.~Semkin was supported in part by the Academy of Finland.

%\ifCLASSOPTIONcaptionsoff
%  \newpage
%\fi

\bibliographystyle{IEEEtran}
\bibliography{bibl}

\end{document}